\documentclass[preprint2]{aastex}
\usepackage{rotating}
\usepackage{graphicx}
\DeclareGraphicsExtensions{.ps} 

\shorttitle{\indent \def Temporal evolution of chromospheric evaporation} \shortauthors{Tian et al.}

\begin{document}

\title{Temporal evolution of chromospheric evaporation: case studies of the M1.1 flare on 2014 September 6 and X1.6 flare on 2014 September 10}

\author{Hui Tian\altaffilmark{1}, Peter R. Young\altaffilmark{2}, Katharine K. Reeves\altaffilmark{1}, Bin Chen\altaffilmark{1}, Wei Liu\altaffilmark{3,4}, Sean McKillop\altaffilmark{1}}
\altaffiltext{1}{Harvard-Smithsonian Center for Astrophysics, 60 Garden Street, Cambridge, MA 02138, USA; hui.tian@cfa.harvard.edu}
\altaffiltext{2}{College of Science, George Mason University, Fairfax, VA 22030, USA} \altaffiltext{3}{Lockheed Martin Solar and Astrophysics
Laboratory, Building 252, 3251 Hanover Street, Palo Alto, CA 94305, USA} \altaffiltext{4}{W. W. Hansen Experimental Physics Laboratory, Stanford
University, Stanford, CA 94305, USA}

\begin{abstract}
With observations from the Interface Region Imaging Spectrograph (IRIS), we track the complete evolution of $\sim$11 MK evaporation flows in an M1.1 flare on 2014 September 6 and an X1.6 flare on 2014 September 10. These hot flows, as indicated by the blueshifted Fe~{\sc{xxi}}~1354.08\AA{}~line, evolve smoothly with a velocity decreasing exponentially from $\sim$200~km~s$^{-1}$ to almost stationary within a few minutes. We find a good correlation between the flow velocity and energy deposition rate as represented by the hard X-Ray flux observed with the Reuven Ramaty High Energy Solar Spectroscopic Imager (RHESSI), or time derivative of the soft X-Ray flux observed with the Geostationary Operational Environmental Satellites (GOES) and the HINODE X-ray Telescope (XRT), which is in general agreement with models of nonthermal electron heating. The maximum blue shift of Fe~{\sc{xxi}}~appears approximately at the same time as or slightly after the impulsive enhancement of the ultraviolet continuum and the Mg~{\sc{ii}}~2798.8\AA{}~line emission, demonstrating that the evaporation flow is closely related to heating of the lower chromosphere. Finally, while the hot Fe~{\sc{xxi}}~1354.08\AA{} line is entirely blueshifted with no obvious rest component, cool chromospheric and transition region lines like Si~{\sc{iv}}~1402.77\AA{} are often not entirely redshifted but just reveal an obvious red wing enhancement at the ribbons, suggesting that the speed of chromospheric condensation might be larger than previously thought.
\end{abstract}


\section{Introduction}
Solar flares are among the most intense energy release events in the solar atmosphere. In the standard flare model
\citep{Carmichael1964,Sturrock1966,Hirayama1974,Kopp1976}, energy released from magnetic reconnection first heats the plasma and accelerates
particles in the corona. The released energy is then transported along the newly reconnected flare loops towards the chromosphere through either
thermal conduction or nonthermal electrons, which results in an intense and rapid heating of the chromospheric material up to a temperature on
the order of 10 MK. The chromospheric heating is revealed through bright flare ribbons in images of H$\alpha$ and UV continuum. At the same
time the nonthermal electrons can produce hard X-Ray (HXR) emission through bremsstrahlung with the thermal ions. The overpressure associated
with the chromospheric heating drives the heated plasma upward to fill the flare loops. This so-called "chromospheric evaporation process" has
been proposed to explain the soft X-Ray (SXR) emission which usually peaks after the HXR peak \citep[e.g.,][]{Neupert1968}. In the impulsive
phase of some flares the time history of HXR flux has been found to closely match the derivative of SXR light curve. This relationship is called
the Neupert effect and is often quoted as a manifestation of chromospheric evaporation
\citep[e.g.,][]{Hudson1991,Dennis1993,Veronig2005,Liu2006}.

There have been only a few broad-band imaging observations of chromospheric evaporation \citep{Silva1997,Liu2006,Ning2009,Nitta2012,Zhang2013}. These authors found X-Ray sources moving towards the loop top during the impulsive phase of the flares and interpreted them as thermal emission at the evaporation front or nonthermal emission due to decreasing stopping distance from the loop top caused by the increasing density. Most of our knowledge about
chromospheric evaporation is based on spectroscopic observations, probably because spectroscopy is the best way to observe vertical flows in
on-disk observations. The most direct spectral signature of the evaporation flow should be the blue shift of hot emission lines. Indeed, early
observations by the X-Ray spectrometers mounted on the P78-1, the Solar Maximum Mission and the \textit{Yohkoh} spacecraft have detected
blueshifted components of SXR emission lines formed at temperatures of 8--25 MK \citep{Doschek1980,Feldman1980,Antonucci1982,Mariska1993},
indicating the presence of hot and fast (up to $\sim$400~km~s$^{-1}$) plasma upflows. Enhancement in the blue wing of the
Fe~{\sc{xxi}}~1354.08\AA{} line indicating an upflowing component at a speed of $\sim$200~km~s$^{-1}$ was also detected by the Ultraviolet
Spectrometer and Polarimeter (UVSP) on board the Solar Maximum Mission \citep{Mason1986}. In all these observations, there is a dominant
stationary emission component which likely results from the surrounding stationary hot plasma in the flare loops.

Observations from the Coronal Diagnostic Spectrometer \citep[CDS,][]{Harrison1995} on board the Solar and Heliospheric Observatory (SOHO) have
greatly improved our understanding of chromospheric evaporation. For example, \cite{Brosius2003} detected blueshifted line profiles from the
O~{\sc{iii}}, O~{\sc{v}}, Mg~{\sc{x}}~and Fe~{\sc{xix}} ions early during the impulsive rise of a flare. He also found that all lines except the
Fe~{\sc{xix}}~line subsequently exhibited multiple components, indicating downflows resulting from cooling of the heated plasma.
\cite{Harra2005} reported two "ribbons" where the Fe~{\sc{xvi}}~and Fe~{\sc{xix}}~line profiles exhibited blue shifts of a few tens km~s$^{-1}$.
These two "ribbons" were found to separate with time, which was interpreted as a signature of chromospheric evaporation in sequentially
reconnected loops. Blue shifts of some transition region lines have also been found before the impulsive phase of some flares
\citep{Brosius2004,Harra2005}. The early evaporation flow has been proposed by \cite{Harra2005} to support the breakout model of coronal mass
eruption \citep{Antiochos1999}, where magnetic reconnection starts before the filament eruption. \cite{Milligan2006a} and \cite{Milligan2006b}
presented examples of gentle and explosive evaporation \citep[e.g.,][]{Fisher1985b} in the impulsive phase of flares, respectively. In the
former case, heating by relatively low flux of nonthermal electrons leads to evaporated plasma upflows at several tens km~s$^{-1}$ with no
associated downflows. In the latter case, heating by relatively high flux of electrons leads to an overpressure which results in both an upward
expansion of the hot plasma at several hundred km~s$^{-1}$ ($\sim$230 km~s$^{-1}$ blue shift of Fe~{\sc{xix}}~592.23\AA{}) and a downward
condensation of the cooler material at several tens km~s$^{-1}$ ($\sim$40 km~s$^{-1}$ red shift of He~{\sc{i}}~584.33\AA{} and
O~{\sc{v}}~629.73\AA{}). \cite{DelZanna2006} also found that the Fe~{\sc{xix}}~592.23\AA{}~line was largely blueshifted by $\sim$140 km~s$^{-1}$
in an explosive evaporation. The Fe~{\sc{xix}}~blue shift was found to be $\sim$80 km~s$^{-1}$ in two flares studied by \cite{Falchi2006} and \cite{Raftery2009}. The Fe~{\sc{xix}}~592.23\AA{}~line appears to be entirely blueshifted in many of these CDS observation. However,
as pointed by \cite{Young2013}, this is likely because the low spectral resolution of CDS blurs the multiple components of the line. Indeed,
Fe~{\sc{xix}}~592.23\AA{}~line profiles with both a nearly stationary component and a highly blueshifted (by $\sim$200 km~s$^{-1}$) component
were identified in a few flares \citep{Teriaca2003,Teriaca2006,Milligan2006b}. Coronal upflows at a speed of several tens of km~s$^{-1}$ have also been found by CDS in the late gradual phase of a flare \citep{Czaykowska1999}. These upflows have been suggested to result from chromospheric evaporation driven by thermal conduction \citep{Czaykowska2001}.

With observations of the EUV imaging spectrometer \citep[EIS,][]{Culhane2007}, our knowledge about chromospheric evaporation has been further
enhanced in the past 6 years. EIS has a better coverage of the coronal temperatures over CDS and allows study of the temperature dependent behavior of the evaporation flows. Although different patterns of evaporation flows have been found in different flares and even at different pixels within
the same ribbon, two surprising results appear to be reported for many flares. First, profiles of hot lines from Fe~{\sc{xxiii}}~(formed at
$\sim$14 MK) and Fe~{\sc{xxiv}}~(formed at $\sim$18 MK) usually exhibit a blueshifted component or blue wing enhancement besides a nearly
stationary component, \citep[e.g.,][]{Milligan2009,Young2013}, which is not consistent with single flare loop models
\citep[e.g.,][]{Emslie1987,Liu2009} and may be explained by an ensemble of sequential heating of many small-scale threads
\citep[e.g.,][]{Warren2005,Reeves2007}. This result suggests that the spatial resolution of EIS ($\sim$3$^{\prime\prime}$) is insufficient to
separate the evaporation flow from the ambient stationary hot plasma. The velocities associated with the blueshifted components can be as high
as $\sim$400 km~s$^{-1}$. In some flares blueshifted components are even present in profiles of warm lines such as Fe~{\sc{xiii}},
Fe~{\sc{xiv}}, Fe~{\sc{xv}}~and Fe~{\sc{xvi}} \citep[formed at 1.5--3 MK,][]{Watanabe2010,DelZanna2011,Li2011,Graham2011}. The velocities
inferred from the profiles of these warm lines are mostly less than 150 km~s$^{-1}$. In most cases the stationary component is dominant,
although a dominant blueshifted component has also been reported in a few flares \citep{Watanabe2010,Li2011,Young2013}. We note, however,
completely blueshifted Fe~{\sc{xxiii}}~line profiles were identified occasionally in a few flares \citep{Brosius2013, Doschek2013}. Second, in explosive
evaporations the temperature at which the Doppler shift turns from red to blue appears to be much higher than those predicted by some models.
For example, the model of \cite{Fisher1985a} predicts a condensation of plasma in the chromosphere. However, EIS observations often show a few
tens km~s$^{-1}$ redshift of emission lines formed at much higher temperatures. For example, \cite{Milligan2009} and \cite{Young2013} found that
the transition occurs in the temperature range of 1.5-2.0 MK and 1.1-1.5 MK, respectively. A high transition temperature of $\sim$0.8 MK was
reported even in a small B1.4 flare \citep{Chen2010}. A recent model of \cite{Liu2009} shows that continuous energy deposition throughout the
impulsive phase can heat the underlying chromosphere up to 2 MK, which may explain the observed red shifts of some coronal lines. Besides the
Doppler shift, the nonthermal line broadening associated with the evaporation flow has also been investigated using EIS data. Significant
nonthermal broadening ($\sim$100 km~s$^{-1}$, FWHM) has been found for the hot Fe~{\sc{xxiii}}~and Fe~{\sc{xxiv}}~lines
\citep{Milligan2011,Young2013}. \cite{Milligan2011} also found a positive correlation between the blue shift and nonthermal width for the
Fe~{\sc{xv}}~and Fe~{\sc{xvi}}~lines at one loop footpoint.

With the launch of the Interface Region Imaging Spectrograph \citep[IRIS,][]{DePontieu2014}, chromospheric evaporation can now be observed at
unprecedented high spatial ($\sim$0.33$^{\prime\prime}$), spectral ($\sim$3 km~s$^{-1}$) and temporal ($\sim$3 s) resolution. Although there is
only one line, Fe~{\sc{xxi}}~1354.08\AA{} (formation temperature $\sim$11 MK from CHIANTI, \cite{Landi2013}), formed above 2 MK in the IRIS
spectral range, significant new insight about chromospheric evaporation has already been revealed from IRIS observations. The most intriguing
finding is the completely blueshifted Fe~{\sc{xxi}}~1354.08\AA{}~line profiles in the evaporation flows, which suggests that the resolution of
IRIS is high enough to fully resolve the evaporation flows. The reported blue shift of the Fe~{\sc{xxi}}~1354.08\AA{}~line appears to be mostly
around 100 km~s$^{-1}$ or smaller \citep{Young2015,Li2015,Polito2015,Sadykov2015}, although a clear $\sim$260 km~s$^{-1}$ blue shift was
reported in a C1.6 flare \citep{Tian2014}. These subarcsecond-resolution observations often reveal a spatial offset between the continuum
enhancement and the blueshifted Fe~{\sc{xxi}}~emission, which is likely due to the fact that the hot evaporation flows are located higher in the
loop legs. \cite{Polito2015} also found a decrease of both the blue shift and nonthermal line width of Fe~{\sc{xxi}}, although only six data
points were measured due to the 75 s cadence of the raster scans.

Here we report results from IRIS sit-and-stare observations of chromospheric evaporation in two flares. With a cadence of $\sim$9.5 s, we
track the complete evolution of hot ($\sim$11 MK) evaporation flows in the M1.1 flare on 2014 September 6 and the X1.6 flare on 2014 September
10. With simultaneous observations from the Reuven Ramaty High Energy Solar Spectroscopic Imager \citep[RHESSI,][]{Lin2002}, the Geostationary
Operational Environmental Satellites (GOES) and the X-ray Telescope \citep[XRT,][]{Golub2007} on board Hinode, we also investigate the possible
correlation of the flow velocity with the energy deposition rate and chromospheric heating. Our observations reveal new insight into the
chromospheric heating and evaporation processes during flares.

\section{The M1.1 flare on 2014 September 6}
\subsection{Observations and data reduction}

\begin{figure*}
\centering {\includegraphics[width=\textwidth]{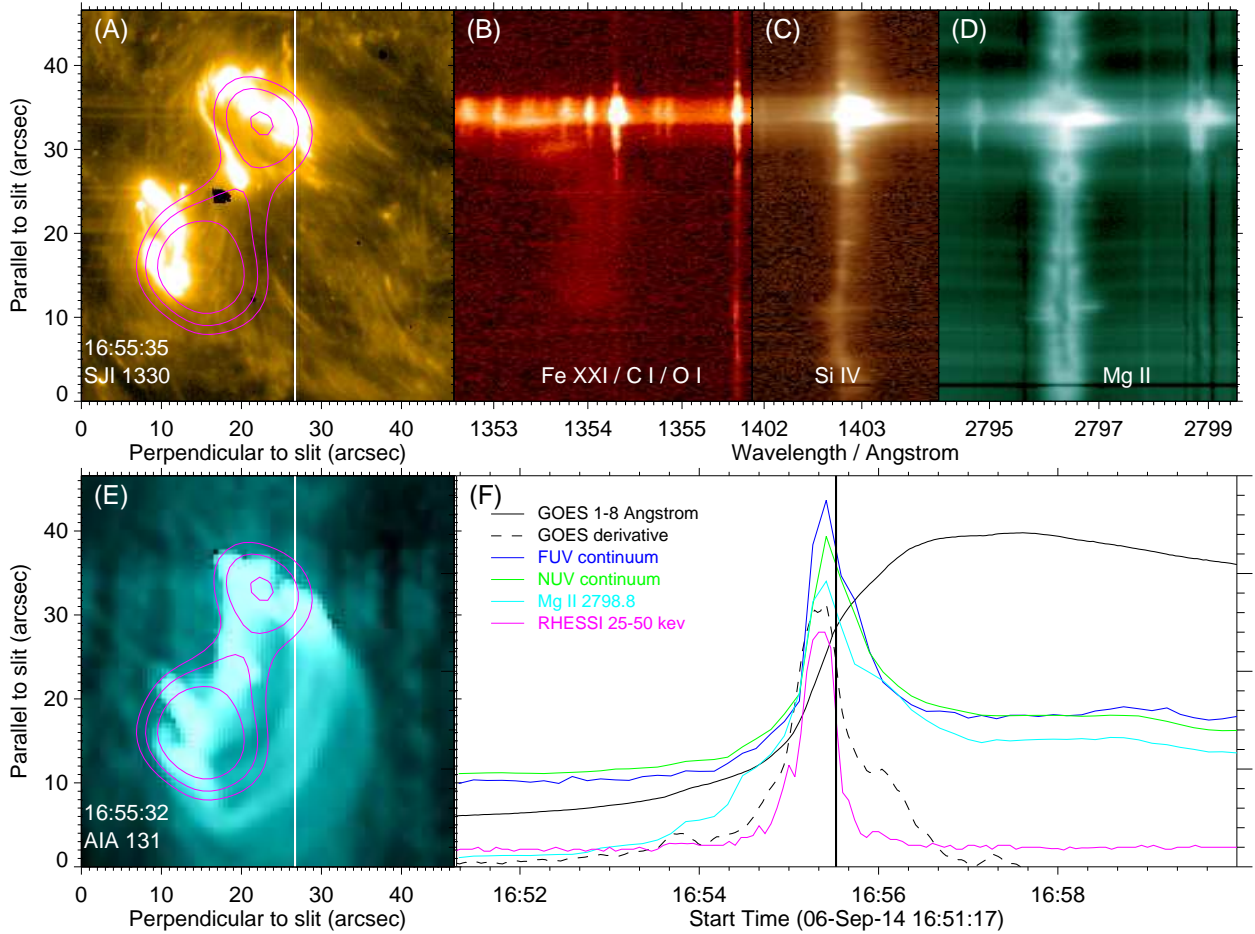}} \caption{ (A): IRIS/SJI~1330\AA{}~image taken at 16:55:35 UT on 2014 September 6.
(B)--(D): IRIS detector images taken at the same time in three spectral windows. (E) SDO/AIA 131\AA{}~image taken at 16:55:32 UT. (F): Light
curves (arbitrarily scaled) of the GOES 1-8{\AA} flux and its time derivative, continua and Mg~{\sc{ii}}~2798.8\AA{}~intensities averaged over
the slit positions of 30$^{\prime\prime}$--39$^{\prime\prime}$, and RHESSI 25--50 keV flux. In panels (A) and (E), the slit location is indicated by the white vertical line, and contours of RHESSI 25--50~keV X-Ray flux are overplotted on the images. The vertical line in panel (F) indicates the observation time of the images. Two online movies m1.mp4
and m2.mp4 are associated with this figure. } \label{fig.1}
\end{figure*}

An M1.1 flare occurred in NOAA AR12157 at about 17:00 UT on 2014 September 6. Two peaks are present in the GOES SXR light curves, suggesting two
episodes of energy release in this flare. Here we focus on the first episode from roughly 16:51 UT to 17:00 UT, when the plasma heating process
was clearly observed by the IRIS slit at one footpoint of the flare loops. In this observation IRIS pointed at (--732$^{\prime\prime}$,
--299$^{\prime\prime}$), with a roll angle of 45 degrees. The slit was fixed at the same location on the Sun throughout the whole observing
period (sit-and-stare, 11:23 UT -- 19:30 UT). The spatial pixel size was 0.166$^{\prime\prime}$. The spectral dispersion was $\sim$0.026 \AA{} per pixel in the far
ultraviolet (FUV) wavelength bands (FUV spectrally summed by 2) and $\sim$0.025 \AA{} per pixel in the near ultraviolet (NUV) wavelength band.
The cadence of the spectral observation was $\sim$9.5 seconds, with an exposure time of 8 seconds. Slit-jaw images (SJI) were taken at a cadence
of $\sim$19 seconds in each of the 1400\AA{}~and 1330\AA{}~passbands, and the exposure time decreased from 8 seconds to 2.4 seconds after
16:55:15 UT. Dark current subtraction, flat field correction, geometrical and orbital variation corrections have been applied in the level 2 data used here. Several relatively strong neutral lines have been used for absolute wavelength calibration, e.g., O~{\sc{i}}~1355.5977\AA{},
S~{\sc{i}}~1401.5136\AA{}~and Mn~{\sc{i}}~2801.908\AA{}. These lines were assumed to be at rest outside the flare regions. The fiducial marks were used to achieve a coalignment of the two SJI passbands and different spectral windows. Figure~\ref{fig.1}(A)-(D) shows the IRIS SJI 1330\AA{}~image and spectral images in three wavelength windows at 16:55:35 UT.

The 131\AA{}~passband of the Atmospheric Imaging Assembly \citep[AIA,][]{Lemen2012} on board the Solar Dynamics Observatory
\citep[SDO,][]{Pesnell2012} samples emission mainly from Fe~{\sc{xxi}}~in flares \citep[e.g.,][]{ODwyer2010}. We use images of this
passband for the examination of the flare loop morphology (Figure~\ref{fig.1}(E)). We notice that early during the impulsive phase this passband is likely dominated by Fe~{\sc{viii}}~lines \citep{Boerner2012,Brosius2012}. We rotated the AIA images by 45 degrees to match the IRIS
spacecraft orientation. We coaligned the AIA 1600\AA{}~(mainly FUV continuum and C~{\sc{iv}}) and IRIS 1330\AA{}~(mainly FUV continuum and
C~{\sc{ii}}) images. AIA 131\AA{}~images were then automatically aligned with the IRIS images since AIA images in different passbands are
automatically coaligned after applying the standard SolarSoft (SSW) routine aia\_prep.pro. The cadence of the AIA 131\AA{}~images is 12 seconds.

RHESSI data is also available for this flare. The RHESSI 25--50~keV image reconstructed with the CLEAN algorithm using detectors 2--8 from
16:55:00 UT to 16:55:32 UT shows two HXR sources, which are clearly located at the two footpoints of the flare loops (Figure~\ref{fig.1}(E)).
Note that the RHESSI image has also been rotated by 45 degrees.

We present the time sequence of the IRIS observation (including SJI 1330\AA{} images and three spectral windows) and the corresponding AIA
131\AA{}~images in an online movie (m1.mp4). The GOES SXR and RHESSI HXR light curves are shown in the movie. The RHESSI 25-50
keV HXR and the time derivative of GOES 1-8{\AA}~SXR show a reasonably good match, indicating the existence of the Neupert effect. We also define the NUV and FUV continua by averaging the IRIS spectra over slit
positions of 30$^{\prime\prime}$--39$^{\prime\prime}$ and wavelength ranges of 1355.08\AA{}--1355.43\AA{}~and 2794.10\AA{}--2794.45\AA{},
respectively. The time history of the Mg~{\sc{ii}}~2798.8\AA{}~intensity (integrated over 2798.65\AA{}--2799.00\AA{}, NUV continuum subtracted)
was also calculated over the same slit positions. The continua and Mg~{\sc{ii}}~light curves appear to be similar. Figure~\ref{fig.1} is a
snapshot of this movie. Movie m2.mp4 shows the time evolution of the Fe~{\sc{xxi}}~1354.08\AA{}~and Si~{\sc{iv}}~1402.77\AA{}~spectra at the
full-cadence ($\sim$9.5 s) in a smaller section of the slit. Note that different images have been scaled differently in this movie to reveal a
better contrast between the Fe~{\sc{xxi}}~emission and the other emission features.

\subsection{Results and discussion}

Figure~\ref{fig.1} shows that the IRIS slit crosses the edge of a ribbon, where one footpoint of the flare loops is anchored. The higher part of
the loops is also covered by the slit. From the online movie m1.mp4 we can see that discernible Fe~{\sc{xxi}}~1354.08\AA{}~emission first
appears at about 16:51:57 UT around the slit position of 28$^{\prime\prime}$. About one minute later, stronger and entirely blueshifted
Fe~{\sc{xxi}}~emission appears around slit position 30$^{\prime\prime}$, where a localized brightening can be identified in both the
1330\AA{}~and 131\AA{}~images. The entirely blueshifted Fe~{\sc{xxi}}~line without a stationary component has been recently reported in several
other flares \citep{Young2015,Li2015,Polito2015,Sadykov2015,Tian2014}, and is in agreement with hydrodynamic simulations of single flare loop
\cite[e.g.,][]{Emslie1987,Liu2009}. Enhancement of both the FUV and NUV continua can be seen at slit positions
30$^{\prime\prime}$--39$^{\prime\prime}$ after 16:53:51 UT, $\sim$2 minutes after the first Fe~{\sc{xxi}}~1354.08\AA{}~signature. During the
impulsive phase of the flare, intensities of the 1330\AA{}~and 131\AA{}~passbands, the transition region and chromospheric emission lines, and
the FUV and NUV continua all increase dramatically. These intensities reach their maxima at around 16:55:25 UT, almost the same as the peak
times of the RHESSI HXR and GOES derivative. The 131\AA{}~emission is likely dominated by Fe~{\sc{viii}}~lines early in the impulsive phase \citep{Boerner2012,Brosius2012}. In the impulsive phase of the flare, the Fe~{\sc{xxi}}~emission is found mainly at or close to the
ribbon where the continua are enhanced, and the line is clearly blue shifted. A visual inspection of the movie also suggests that the
Fe~{\sc{xxi}}~blue shift reaches the maximum around the peak times of the RHESSI HXR and GOES derivative. The flare loops have little
Fe~{\sc{xxi}}~1354.08\AA{}~emission in the early stage of the impulsive phase. However, more and more Fe~{\sc{xxi}}~emission appears to move
upward from the loop footpoint to higher part in the later stage of the impulsive phase, and eventually fills the whole flare loops around the
peak time of the GOES flux. This process provides strong support to the argument that the SXR emission in flare loops is produced through
chromospheric evaporation, although we can not rule out the possibility that some of the coronal plasma might be heated in situ
\citep[e.g.,][]{Liu2013}. The Fe~{\sc{xxi}}~emission from these flare loops is very strong and shows almost no Doppler shift.

\begin{sidewaysfigure*}
\centering {\includegraphics[width=\textwidth]{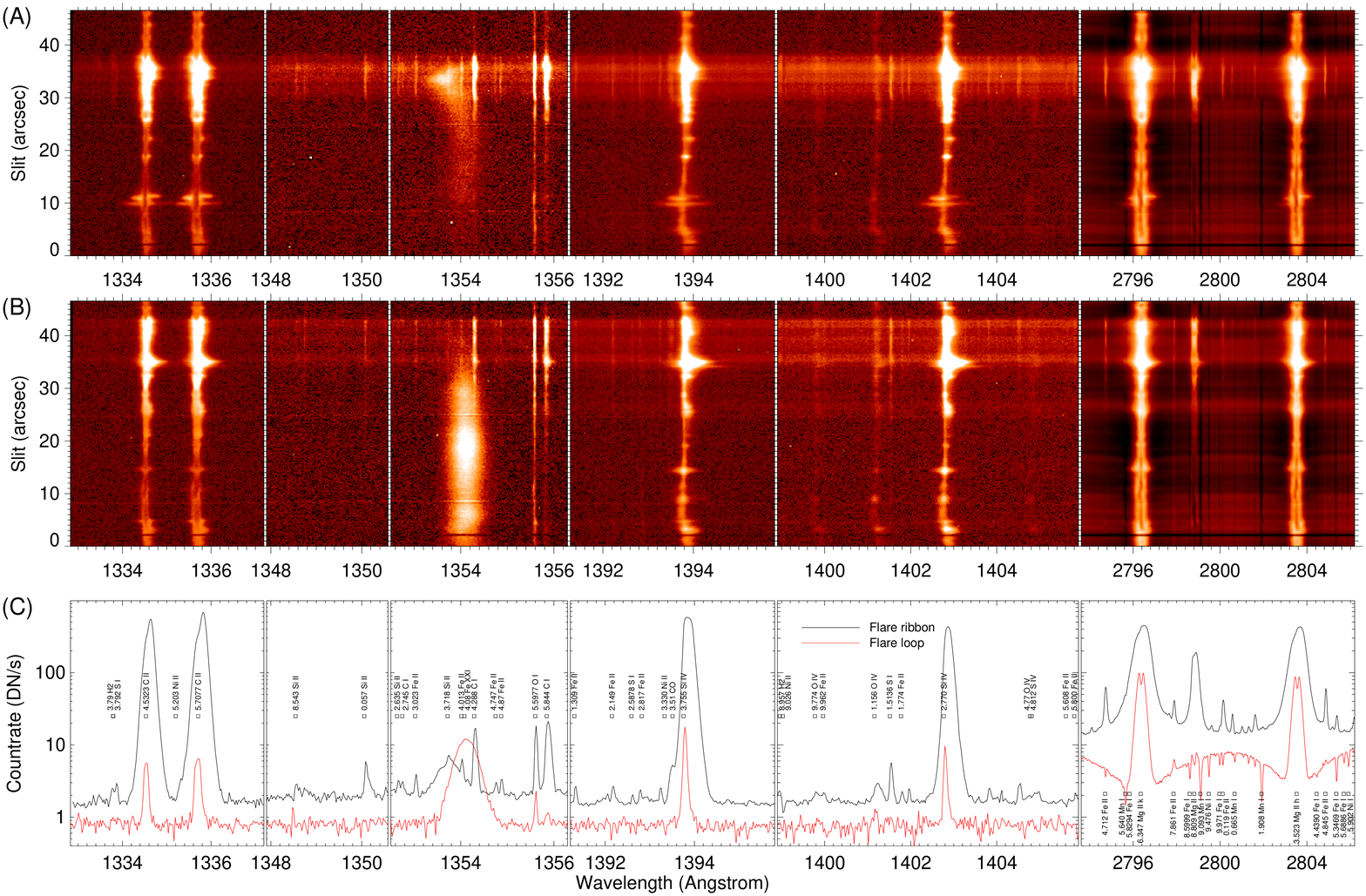}} \caption{Typical IRIS spectra in the M1.1 flare on 2014 September 6. (A) Detector images
taken at 16:56:13 UT. (B) Detector images taken at 17:08:59 UT. (C) Typical spectral line profiles at the flare ribbon (averaged over 10 pixels
around slit position 34.9$^{\prime\prime}$ in panel (A)) and flare loop (averaged over 10 pixels around slit position 16.6$^{\prime\prime}$ in
panel (B)). The flare ribbon spectra in the two Si~{\sc{iv}}~spectral windows have been divided by 3 for the purpose of illustration. }
\label{fig.2}
\end{sidewaysfigure*}

Figure~\ref{fig.2} presents typical spectra at the flare ribbons and flare loops in this flare. The ribbon spectra and loop spectra appear to be
largely different. First, almost all emission lines as well as the FUV and NUV continua are much stronger at the ribbons than in the loops. The
Fe~{\sc{xxi}}~1354.08\AA{}~line, however, is often stronger in the loops. This is because hot plasma accumulates in the flare loops as the
evaporation proceeds. Second, many narrow lines, which are normally invisible, now appear in the ribbon spectra. Most of these lines are singly
ionized or neutral lines, and some of them have been identified with previous instruments
\citep{Cohen1981,Sandlin1986,Feldman1997,Ekberg2003,Curdt2001,Curdt2004,Tian2009}. Some lines and their vacuum wavelengths
\citep{Kelly1987,Kelly1979} have been marked in Figures~\ref{fig.2}. Third, the NUV spectra at the flare ribbons are dominated by emission
lines \citep[see also][]{Cheng2015b}. This is the case even at longer wavelengths around 2832\AA{} (not shown here), indicating photospheric heating through possibly radiative
backwarming \citep[e.g.,][]{Xu2004}. The Mg~{\sc{ii}}~2798.809\AA{}~line, which is actually a blend of two Mg~{\sc{ii}}~lines at
2798.754\AA{}~and 2798.822\AA{}, is the strongest emission line in the NUV spectrum except for the Mg~{\sc{ii}}~k and h lines. These two lines
(Mg~{\sc{ii}}~2798.8\AA{}~line thereafter) are of particular interest as they can serve as a good diagnostic of heating in the lower
chromosphere \citep{Pereira2015}. Away from the flare ribbons, the NUV spectra are similar to quiet Sun spectra with all lines except the
Mg~{\sc{ii}}~k and h in absorption. Fourth, transition region lines such as C~{\sc{ii}}~and Si~{\sc{iv}} are often obviously broader and
redshifted at the ribbons.

\begin{figure*}
\centering {\includegraphics[width=\textwidth]{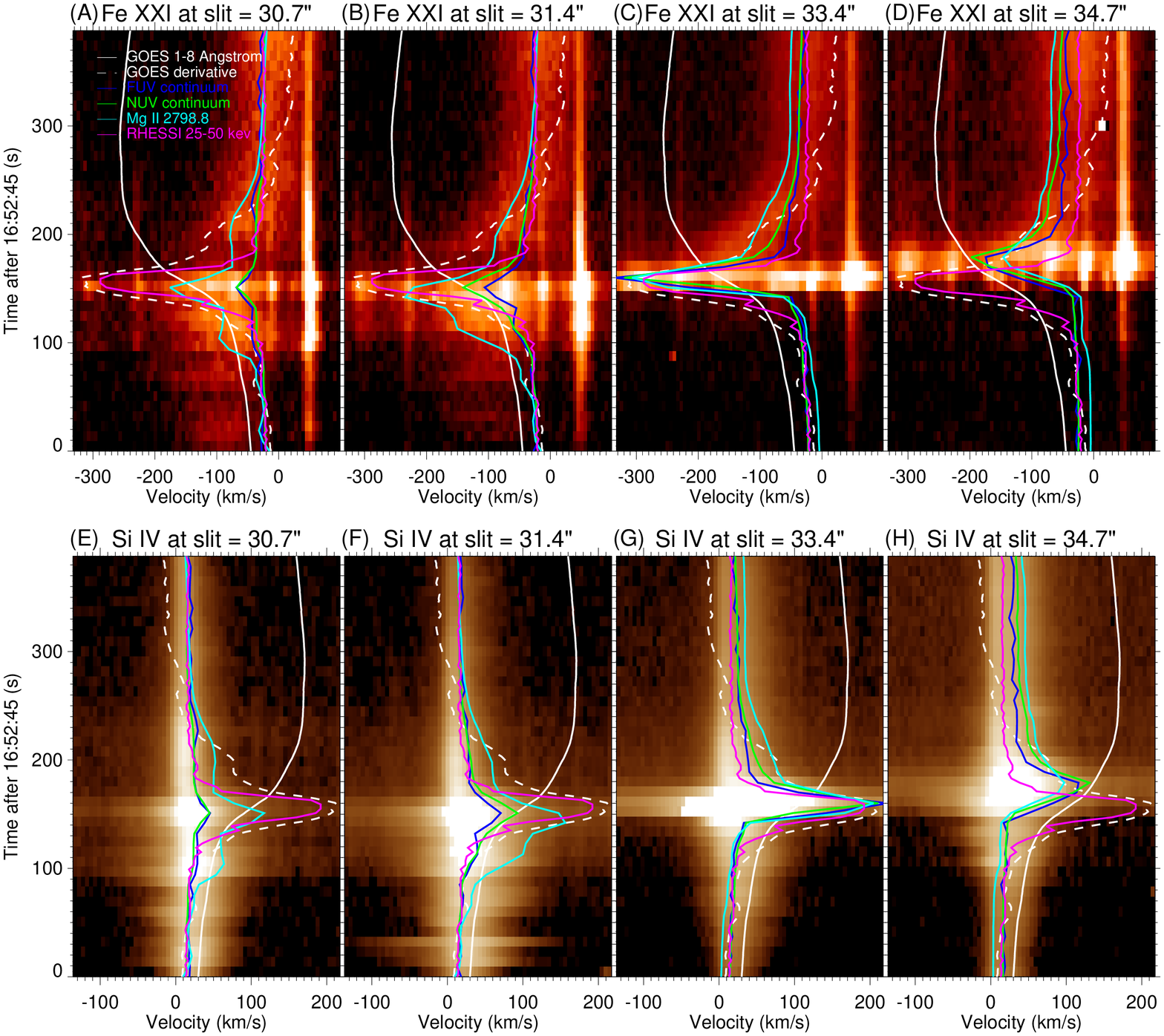}} \caption{ Time evolution of the spectral line profiles of Fe~{\sc{xxi}}~1354.08\AA{}~and
Si~{\sc{iv}}~1402.77\AA{} at four different positions on the slit in the 2014 September 6 observation. The zero velocity has been set to the rest wavelength of the Fe~{\sc{xxi}}~(upper panels) or Si~{\sc{iv}}~(lower panels) line. Negative and positive velocities represent wavelengths shorter (blue shift) and longer (red shift) than the rest wavelength, respectively. The RHESSI 25--50 keV flux (purple), GOES 1-8{\AA} flux (white solid) and its time derivative (white dashed) are overplotted. Light curves of the UV continua (blue for FUV and green for NUV) and Mg~{\sc{ii}}~2798.8\AA{}~(cyan) intensities at the corresponding slit positions are also overplotted.} \label{fig.3}
\end{figure*}

In the following we focus on the flare ribbon region (loop footpoint) crossed by the IRIS slit. Figure~\ref{fig.3} shows the wavelength-time
plots for Fe~{\sc{xxi}}~1354.08\AA{}~and Si~{\sc{iv}}~1402.77\AA{}~at four different positions of the ribbon. At slit positions
30.7$^{\prime\prime}$ and 31.4$^{\prime\prime}$, Fe~{\sc{xxi}}~emission with a blue shift of $\sim$100 km~s$^{-1}$ are clearly present before
the FUV continuum enhancement. This Fe~{\sc{xxi}}~emission is probably from a loop with a footpoint not under the slit. At slit positions
33.4$^{\prime\prime}$ and 34.7$^{\prime\prime}$, there is essentially no Fe~{\sc{xxi}}~emission prior to the continuum enhancement. The maximum
continuum enhancement occurs at or slightly after the peak time of the RHESSI HXR and GOES derivative (for the different spatial scales from which the X-Ray and UV curves are derived, we refer to a discussion later in this section). Since the UV continuum is a marker of
chromospheric heating, this timing suggests that nonthermal electrons deposit energy in the chromosphere and heat the plasma there. Following
the maximum continuum enhancement, we see the appearance of largely blueshifted Fe~{\sc{xxi}}, especially at slit position
33.4$^{\prime\prime}$. The blue shift of Fe~{\sc{xxi}}~smoothly decreases to nearly zero within $\sim$3 minutes, whereas the HXR and UV
continuum appear to decay even faster.

Although a few profiles of Si~{\sc{iv}}~1402.77\AA{}~are saturated around the times of maximum continuum enhancement, it is still clear that the
intensities of Si~{\sc{iv}}, UV continua and Mg~{\sc{ii}}~2798.8\AA{} all enhance at the same time and evolve in a very similar way, which suggests that the Si~{\sc{iv}} emission is responsible for the enhanced emission in lower layers and thus supports the idea that the FUV continuum enhancement at the flare ribbons is driven by back-warming (heating from the downward emission component) from the strong
C~{\sc{ii}} and Si~{\sc{iv}} lines \citep{Doyle1992}. Another interesting feature of the Si~{\sc{iv}} line profiles is the obvious asymmetry of the
line profiles. Red wing asymmetries have been previously reported in the H$\alpha$ line profiles acquired through some ground based observations
of flares \citep[e.g.,][]{Ding1995}, whereas observations by CDS and EIS usually reveal a nearly symmetric profiles of the chromospheric and
transition region lines during flares. The clear enhancement at the red wing suggests the presence of downflows with speeds of generally a few
tens of km~s$^{-1}$. Some of these downflows, e.g., those at the time of maximum continuum enhancement, most likely indicate the "chromospheric
condensation" in explosive evaporation \citep{Fisher1985a}. The red wing enhancement can extend to more than 100 km~s$^{-1}$, suggestive of downflows at speeds greater than previously thought. Note that in many previous observations the downflow speeds were estimated based on a simple Gaussian fit to the line
profiles of cool lines. The single Gaussian fit tends to underestimate the magnitude of red shift if strong red wing asymmetry is present. The
red wing enhancement experiences a fast decay before reaching a stable level. The initial decay appears to be faster than the
Fe~{\sc{xxi}}~blue shift, consistent with the model of \cite{Fisher1989} which predicts a chromospheric condensation time of $\sim$1 minute. We
find that the red wing asymmetry is present even after the impulsive phase. These asymmetries indicate downflows with speeds of $\sim$30
km~s$^{-1}$, which may be the "warm rain" resulting from cooling of the heated plasma, as discussed by \cite{Brosius2003}.

\begin{figure*}
\centering {\includegraphics[width=0.8\textwidth]{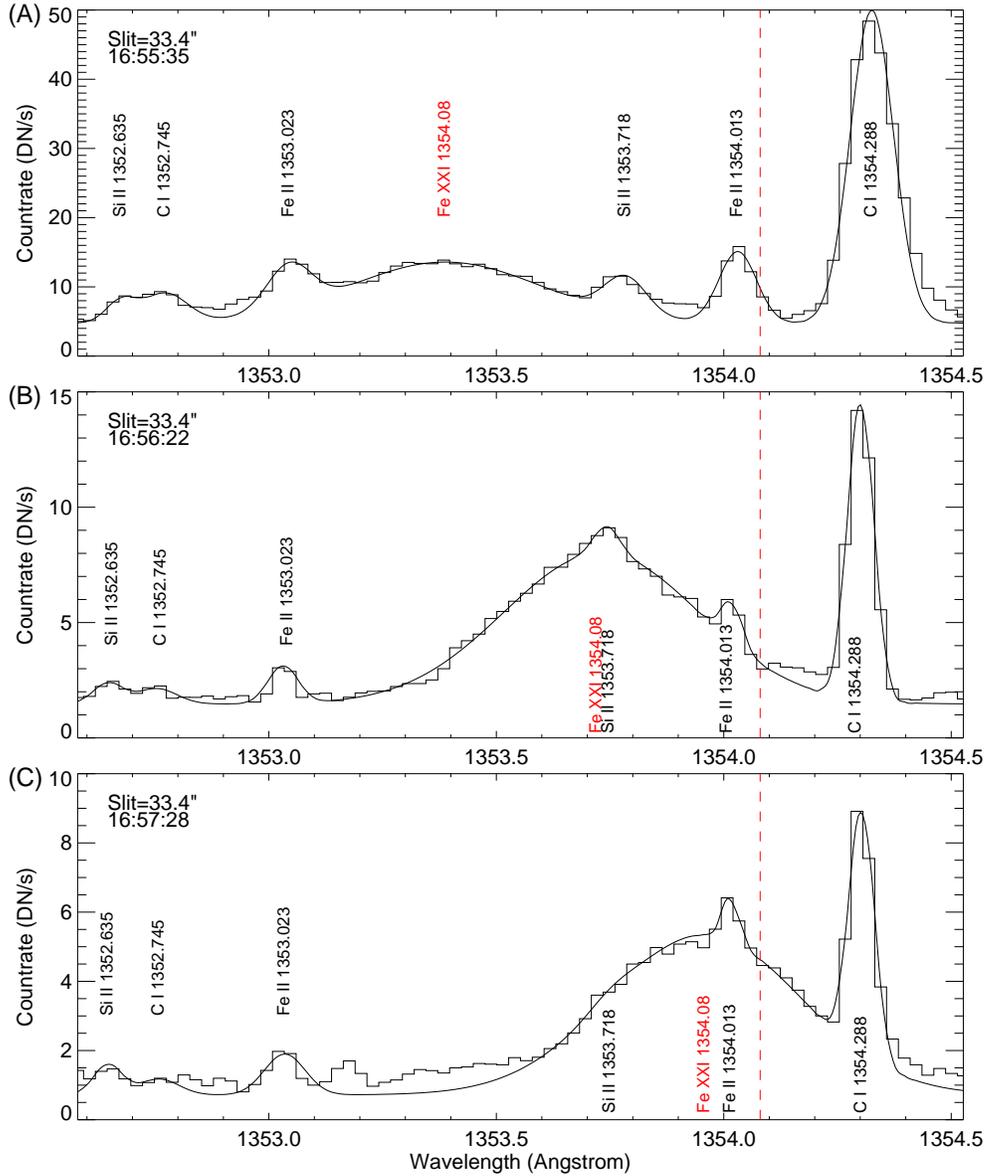}} \caption{ Three examples of multi-Gaussian fit to Fe~{\sc{xxi}}~1354.08\AA{}~and
several lines nearby in the 2014 September 6 observation. The observed and fitted line profiles are shown as the histograms and smooth curves,
respectively. The dashed line indicates the rest wavelength of Fe~{\sc{xxi}}~1354.08\AA{}. } \label{fig.4}
\end{figure*}

\begin{figure*}
\centering {\includegraphics[width=0.8\textwidth]{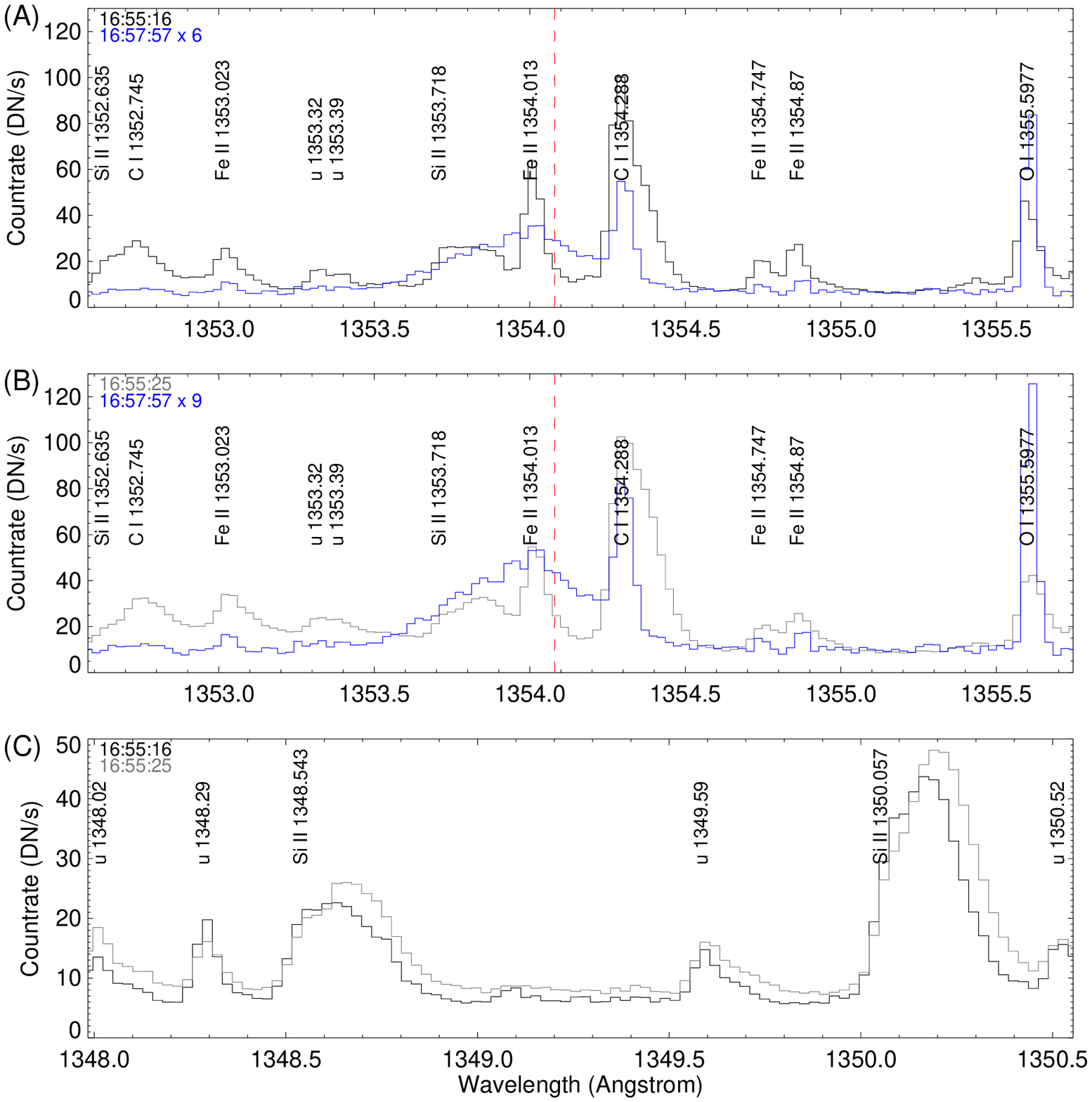}} \caption{ Line profiles of the 1354\AA{}~(A and B) and 1349\AA{}~(C) spectral windows
at 16:55:16 UT and 16:55:25 UT on 2014 September 6, at slit position 33.4$^{\prime\prime}$. As a comparison the line profile at 16:57:57 UT is
re-scaled and overplotted in panels (A) and (B). The dashed line indicates the rest wavelength of Fe~{\sc{xxi}}~1354.08\AA{}. Several previously
unidentified lines are marked as the 'u' followed by the approximate wavelengths.} \label{fig.5}
\end{figure*}

We derive the Fe~{\sc{xxi}}~1354.08\AA{}~line parameters (peak intensity, Doppler shift and line width) in the evaporation flow through
multi-Gaussian fit to the Fe~{\sc{xxi}}~1354.08\AA{}~line and several lines nearby. In the spectral range of 1352.6\AA{}--1354.5\AA{}, the other
major emission lines besides Fe~{\sc{xxi}}~1354.08\AA{}~are the following: Si~{\sc{ii}}~1352.635\AA{}~and 1353.718\AA{},
C~{\sc{i}}~1352.745\AA{}~and 1354.288\AA{}, Fe~{\sc{ii}}~1353.023\AA{}~and 1354.013\AA{}. These six lines are all emitted from neutrals or
singly ionized ions, thus are usually much narrower compared to the Fe~{\sc{xxi}}~1354.08\AA{}~line. These lines also have zero or small Doppler
shift (often red shift). The most prominent and blueshifted Fe~{\sc{xxi}}~emission is found around slit position 33.4$^{\prime\prime}$ after the HXR peak. Before the HXR peak the Fe~{\sc{xxi}}~emission is found slightly to the south of this position (first around slit position 30.7$^{\prime\prime}$ and then 31.4$^{\prime\prime}$, see Figure~\ref{fig.3}), which might be related to some pre-flare activity or some energy release early in the flare. We thus choose three slightly different slit positions during three time periods: slit position 30.7$^{\prime\prime}$ during 16:52:45-16:54:38 UT, 31.4$^{\prime\prime}$ during 16:54:48-16:55:07 UT, and 33.4$^{\prime\prime}$ after 16:55:16 UT. For each exposure, we average the line profiles over 5 pixels around the corresponding slit position to improve the signal to noise ratio. A seven-component Gaussian function has been fitted to each of these spatially averaged line profiles. Figure~\ref{fig.4} shows three examples of multi-Gaussian fit. In some spectra there appear to be some additional emission peaks (e.g., Figure~\ref{fig.4}(C)), suggesting the presence of some unidentified lines. Most of these lines are also very narrow and often weak in intensity, and their presence should not significantly affect the derived line parameters of Fe~{\sc{xxi}}~1354.08\AA{}.

At first sight the Fe~{\sc{xxi}}~1354.08\AA{}~line seems absent at slit position 33.4$^{\prime\prime}$ from 16:55:16 to 16:55:25 UT.
However, a careful examination of the spectra at 16:55:16 and 16:55:25 UT (Figure~\ref{fig.5}) suggests some extra emission around the short
wavelength edge of the spectral window. To demonstrate this we rescaled the line profile obtained at 16:57:57 UT so that the line free part of
the spectrum from 1355.1\AA{}--1355.4\AA{}~matches those at 16:55:16 and 16:55:25 UT. We see obvious enhancement of the continuum level
between the emission lines around the short wavelength edge of the spectral window at 16:55:16 UT and 16:55:25 UT. Moreover, the
C~{\sc{i}}~1352.745\AA{}~line is obviously weaker than the Fe~{\sc{ii}}~1353.023\AA{}~line in all the selected spectra except the two obtained
at 16:55:16 and 16:55:25 UT. It is likely that this bulk enhancement is due to the highly blueshifted ($\sim$200--300~km~s$^{-1}$)
Fe~{\sc{xxi}}~line. However, we realize that it may also be complicated by the redshifted emission of the cool lines. For instance, the
continuum enhancement around 1353.2\AA{}~may be contributed by the enhanced red wing emission of the Fe~{\sc{ii}}~1353.023\AA{}~line.
\cite{Young2015} noticed that several Si~{\sc{ii}}~lines sometimes show broad and extended red wings. This is also the case in our data. Strong
and broad redshifted emission features are clearly present in the Si~{\sc{ii}}~1353.718\AA{}~line profiles at 16:55:16 and 16:55:25 UT
(Figure~\ref{fig.5}(A)-(B)). Similar features are also found for another two Si~{\sc{ii}}~lines at 1348.543\AA{}~and 1350.057\AA{}
(Figure~\ref{fig.5}(C)), demonstrating that they are not unidentified lines. Thus, the enhanced emission of C~{\sc{i}}~1352.745\AA{}~and the
continuum around 1352.9\AA{} may actually be caused by the broad redshifted component of the Si~{\sc{ii}}~1352.635\AA{}~line. Because of this
complication, we conclude that highly blueshifted Fe~{\sc{xxi}}~line possibly exists but cannot be unambiguously identified at 16:55:16 and
16:55:25 UT.

\begin{figure*}
\centering {\includegraphics[width=0.8\textwidth]{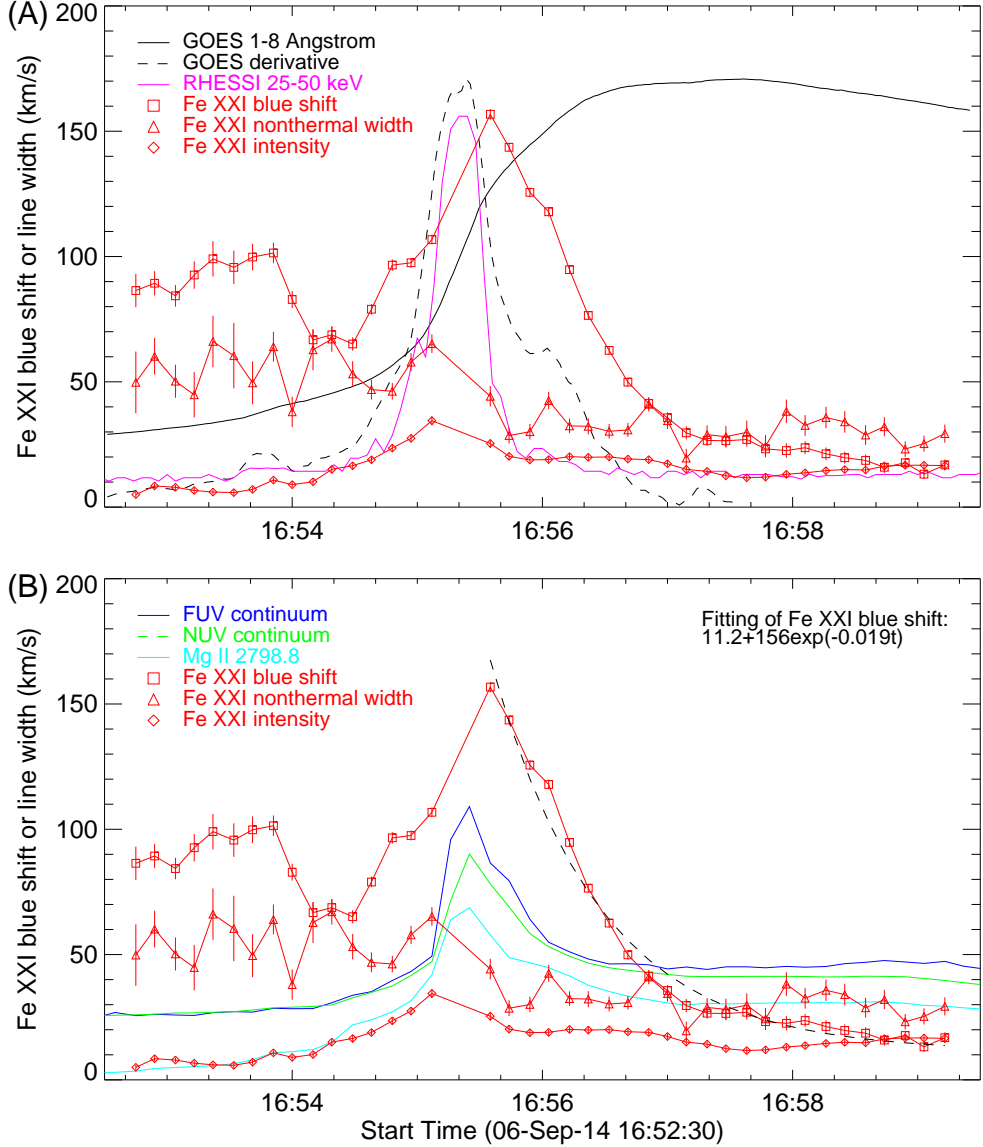}} \caption{ Time history of the line parameters of Fe~{\sc{xxi}}~1354.08\AA{}, the GOES
1-8\AA{} flux and its time derivative, RHESSI 25--50 keV flux, continua and Mg~{\sc{ii}}~2798.8\AA{}~intensities averaged over slit positions of
30$^{\prime\prime}$--39$^{\prime\prime}$, in the 2014 September 6 observation. The error bars represent the 1-$\sigma$ uncertainties from the
multi-Gaussian fitting. Note that the Fe~{\sc{xxi}}~line cannot be unambiguously identified at 16:55:16 UT and 16:55:25 UT, so there are no
measurements of the Fe~{\sc{xxi}}~line parameters at these two times. The black dashed line in panel (B) is the exponential fit of the time
evolution of Fe~{\sc{xxi}}~blue shift. The variable t is the time (second ) after 16:55:35 UT. } \label{fig.6}
\end{figure*}

The derived Fe~{\sc{xxi}}~1354.08\AA{}~line parameters are plotted in Figure~\ref{fig.6}. Our study is among one of the few studies where hot evaporation flows are tracked at a cadence of $\sim$10 s. Using CDS data, the blue shift of Fe~{\sc{xix}}~592.23\AA{} was tracked with a cadence of 9.8 s in two flares \citep{Brosius2003,Brosius2004}. A cadence of 7 s was used in the observation presented by \cite{Brosius2010} but no blueshifts were found for the hot Fe~{\sc{xix}}~line. With EIS observations, \cite{Brosius2013} reported the change of the intensity and Doppler shift of Fe~{\sc{xxiii}}~263.8\AA{} in a C1 flare with a cadence of 11s. We find that the line width values (1/e width) are mostly in the range of 60--90 km~s$^{-1}$. Given a thermal broadening of 57.8 km~s$^{-1}$ (under ionization equilibrium) and an instrumental broadening of $\sim$4 km~s$^{-1}$ \citep[see the discussion in Section S5 of the Supplementary Materials of][]{Tian2014b},
the nonthermal broadening has been found to be in the range of 15--70 km~s$^{-1}$. The line intensity appears to peak around the HXR peak time.
We have to bear in mind that the line parameters are measured at the loop footpoint and that the Fe~{\sc{xxi}}~emitting plasma continuously
moves to higher part of the flare loops. As we explained above, the maximum blue shift of Fe~{\sc{xxi}}~appears in the time range of 16:55:16 UT
-- 16:55:35 UT, roughly the same as the peak times of the RHESSI HXR and the derivative of GOES SXR. The blue shift quickly decreases to
$\sim$20 km~s$^{-1}$ within $\sim$3 minutes, before the start (17:02 UT) of the second episode of energy release in this flare. The uncertainty on the rest wavelength of the Fe~{\sc{xxi}}~line may slightly change the derived Doppler shifts. In our calculation we use 1354.08\AA{}~as the rest wavelength \citep{Sandlin1977}. Using a wavelength of 1354.064\AA{}~given by \cite{Feldman2000} will lower the derived blue shifts by $\sim$3.5 km/s. While using a wavelength of 1354.106\AA{}~found by \cite{Young2015} will increase the derived blue shifts by $\sim$5.8 km/s. The continua and
Mg~{\sc{ii}}~2798.8\AA{}~intensities averaged over the whole covered ribbon (slit positions of 30$^{\prime\prime}$--39$^{\prime\prime}$) reach
their maxima at around 16:55:25 UT, which is almost the same time of the maximum blue shift. Note that the Fe~{\sc{xxi}}~emission comes from
relatively higher part of the loop legs, whilst the continua and Mg~{\sc{ii}}~emission come from lower part of the loop legs. Ideally one should
compare the Doppler shift of Fe~{\sc{xxi}}~and the continua/Mg~{\sc{ii}}~intensities in the same loop. However, it is difficult to do do due
to the line-of-sight superposition effect.

We have to
mention that although the RHESSI HXR and GOES SXR are both integrated over the whole Sun, a comparison of them with the UV emission and 
Fe~{\sc{xxi}}~1354.08\AA{}~line parameters at the ribbon is still meaningful. First, most of the full-Sun RHESSI HXR flux is emitted from the two conjugate footpoints (see Figure~\ref{fig.1}), each located on one of the two ribbons. Connected by the same flaring loop, such HXR footpoints have light curves of the same temporal evolution \citep{Liu2009b}. Therefore, the full-Sun HXR flux is proportional to that of the footpoint crossed by the IRIS slit. The Neupert effect suggests that the GOES derivative is a proxy for the variation of the HXR flux, so it can be compared with the Fe~{\sc{xxi}}~line parameters at the ribbon as well. Second, a recent study clearly reveals that the temporal evolution of evaporation flows at different spatial pixels within the ribbon is strikingly similar \citep{Graham2015}, although we keep in mind that the maximum flow speed might occur at different times at different pixels in some flares.

\begin{figure*}
\centering {\includegraphics[width=0.8\textwidth]{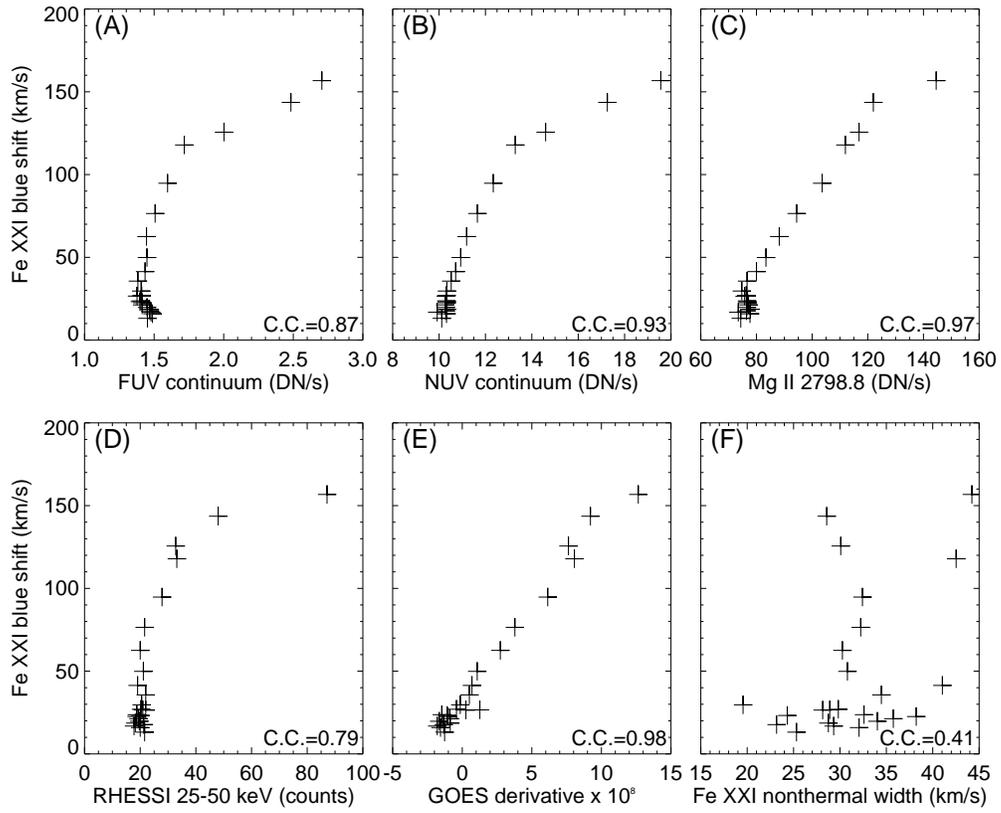}} \caption{ Scatter plots showing the correlation between Fe~{\sc{xxi}}~blue shift and
other parameters in the M1.1 flare on 2014 September 6. The linear Pearson correlation coefficient (C.C.) is marked in each panel. Only data
points measured after 16:55:30 UT are used for the scatter plots. } \label{fig.7}
\end{figure*}

To quantify possible correlation of the Fe~{\sc{xxi}}~blue shift with other parameters, we present various scatter plots in Figure~\ref{fig.7}
for measurements at slit position 33.4$^{\prime\prime}$ after 16:55:30 UT. Larger blue shifts are found when the intensities of
Mg~{\sc{ii}}~2798.8\AA{} and FUV/NUV continuum are larger, supporting the scenario of evaporation flows resulting from heating in the lower
chromosphere.

The Fe~{\sc{xxi}}~blue shift also correlates with the RHESSI HXR and GOES derivative, meaning that the evaporation flow speed depends on the
rate of energy input at the flare ribbon. Perhaps the most intriguing correlation is the one between the blue shift and GOES derivative, which
is nearly linear with a correlation coefficient of 0.98. In the thick-target flare model, most of the accelerated electrons produce HXR through bremsstrahlung in the
chromosphere. The HXR flux is positively correlated with the instantaneous energy deposition rate by the nonthermal electrons
precipitating to the chromosphere. The electron energy deposition may be lost through radiation, or go to the heating of the chromospheric plasma.
Our results suggest that the evaporation flow is enhanced when the energy deposition rate is higher, thus indicating that a significant fraction
of the electron energy goes into chromospheric heating and the resultant evaporation in this flare. The evaporation flow velocity appears to depend
on the heating rate. Recently \cite{Krucker2015} claimed that energy in high-energy electrons ($>$30 keV) is not responsible for chromospheric
evaporation of hot plasma but is lost through radiation in the optical range. Through analysis of an X1 flare, \cite{Heinzel2014} concluded that
the enhanced NUV continuum is the hydrogen recombination Balmer-continuum emission following an impulsive ionization caused by the precipitation
of electrons in the chromosphere, which also suggests the loss of electron energy through radiation. Our
results are not necessarily in contradiction with these previously reported results, as the amount of chromospheric heating and optical
radiation may both increase as the deposited energy increases.

It is interesting that the Fe~{\sc{xxi}}~blue shift is better correlated with the SXR derivative than the HXR and UV continua, although there
may be some uncertainty due to the different spatial scales of integration. We offer possible explanations for this behavior: The blue shift
represents how fast the heated plasma is transported into the flare loops, which may be more directly related to the energy deposition rate into
the flare loops, as represented by the SXR derivative. The HXR and UV continua are also correlated with the energy deposition rate, but they
are, however, produced at the loop foopoints/ribbons. Therefore a positive correlation between them and the Fe~{\sc{xxi}}~blue shift is
expected, but may not be very tight, especially when we consider the nonlinearity in the complex processes involving Coulomb collisions, ionization, free-bound and bremsstrahlung radiation, and hydrodynamic response of the atmosphere. Another possible explanation is that the chromospheric evaporation is mainly driven by thermal conduction rather than direct electron heating \citep[e.g.,][]{Zarro1988,Battaglia2009}, which may lead to a better correlation between the blue shift and SXR derivative.

Although there is a weak trend that smaller blue shifts are associated with smaller nonthermal widths, we do not see a strong correlation between the
blue shift and nonthermal width as reported by \cite{Polito2015} in another flare.

\section{The X1.6 flare on 2014 September 10}
\subsection{Observations and data reduction}

\begin{figure*}
\centering
\begin{minipage}[t]{0.5\textheight}
{\includegraphics[width=\textwidth]{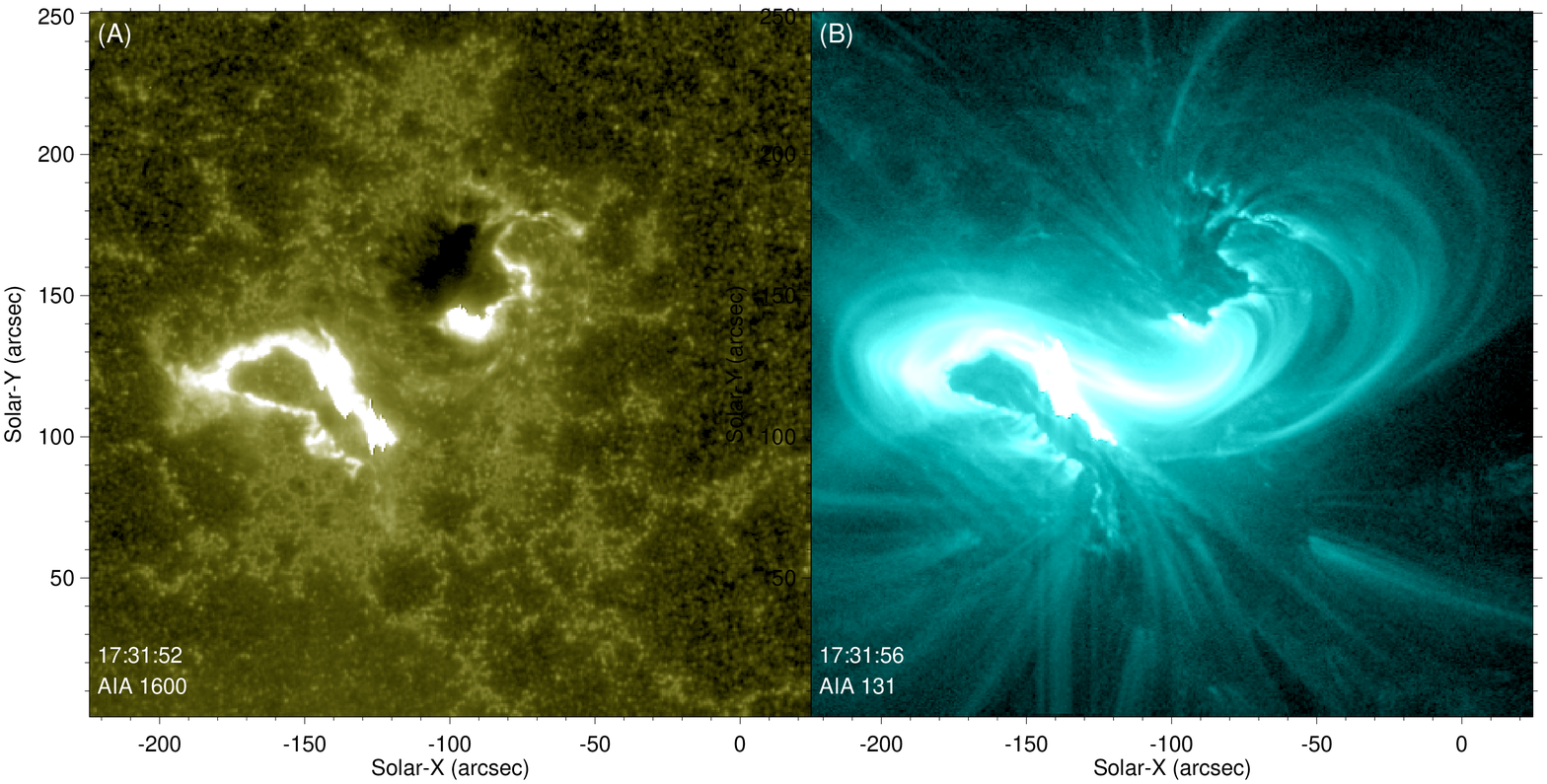}}
\end{minipage}
\begin{minipage}[t]{0.5\textheight}
{\includegraphics[width=\textwidth]{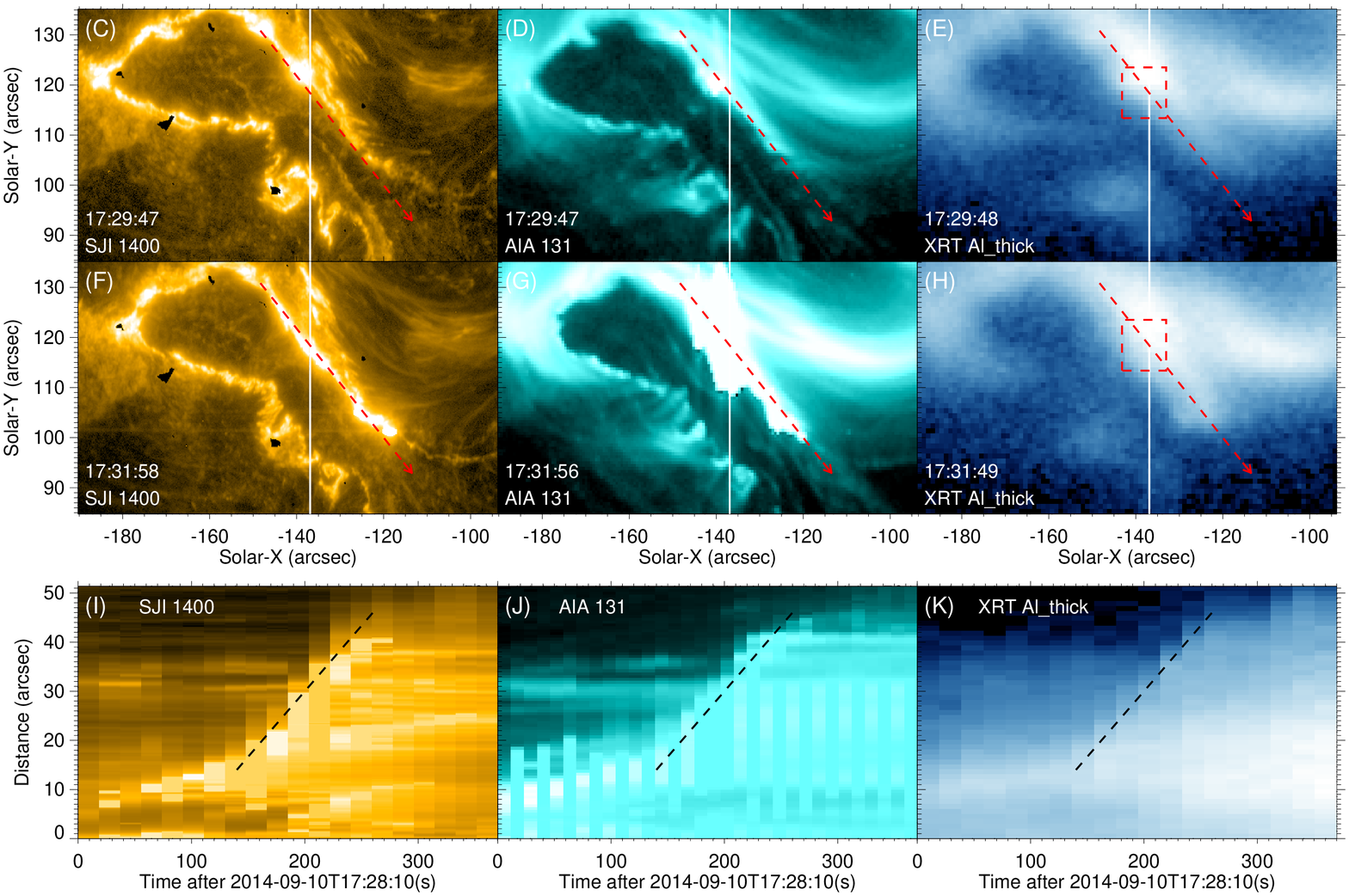}}
\end{minipage}
\begin{minipage}[b]{\textwidth}
\caption{ Context images of the X1.6 flare on 2014 September 10. (A)-(B) Still images of AIA 1600\AA{} and 131\AA{}~from an online movie m3.mp4
showing the evolution of the entire flare region. (C)-(H): IRIS/SJI~1400\AA{}, AIA 131\AA{} and XRT Al\_thick images of the eastern ribbon taken
around 17:29:47 UT and 17:31:56 UT. The red arrow in each panel indicates a fast apparent motion along the ribbon. The vertical line indicates
the slit location. The box shown in panels (E) and (H) marks the region where the XRT intensity is averaged. (I)-(K): Space-time maps for the cut
indicated by the red arrow in panels (C)-(H). The slope of the black line is used to estimate the speed of the apparent motion. An online movie
m4.mp4 shows the evolution of the XRT intensity in which the FOV of Figure~\ref{fig.8}(C)-(H) is indicated by the rectangle. } \label{fig.8}
\end{minipage}
\end{figure*}

\begin{figure*}
\centering {\includegraphics[width=\textwidth]{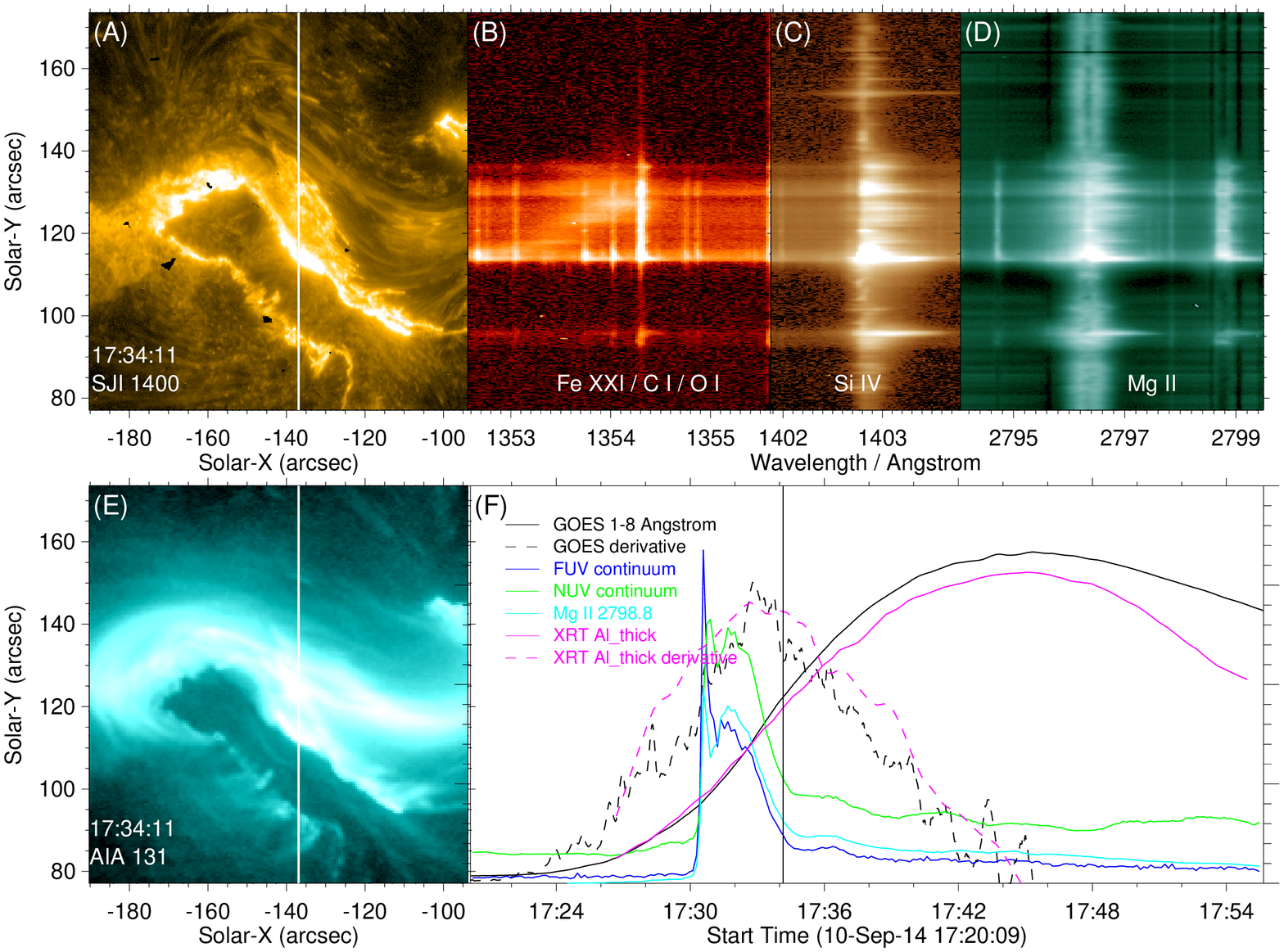}} \caption{ (A): IRIS/SJI~1400\AA{}~image taken at 17:34:11 UT on 2014 September 10.
(B)--(D): IRIS detector images taken at the same time in three spectral windows. (E) SDO/AIA 131\AA{}~image taken at 17:34:11 UT. (F): Light
curves (arbitrarily scaled) of the GOES 1-8\AA{} flux and its time derivative, continua and Mg~{\sc{ii}}~2798.8\AA{}~intensities averaged over 5
pixels around the slit position 117.8$^{\prime\prime}$, XRT Al\_thick intensity integrated over the box shown in Figure~\ref{fig.8}(E) and its
time derivative. In panels (A) and (E), the slit location is indicated by the white vertical line. The vertical line in panel (F) indicates the
observation time of the images. Two online movies m5.mp4 and m6.mp4 are associated with this figure. } \label{fig.9}
\end{figure*}

The second flare we have analyzed is the X1.6 flare peaking at about 17:45 UT on 2014 September 10 from the sigmoidal region in NOAA AR 12158.
The IRIS observation lasted from 11:28 UT to 17:59 UT, which covers the whole impulsive phase and part of the decay phase of the flare. The IRIS slit was oriented in the nominal north-south direction. With a pointing of (--168$^{\prime\prime}$, 125$^{\prime\prime}$), the slit crossed two locations of the eastern ribbon in this flare (Figure~\ref{fig.8}). The observing mode was
also sit-and-stare. The spatial pixel size, FUV spectral pixel size, NUV spectral pixel size, and cadence were 0.166$^{\prime\prime}$,
$\sim$0.026 \AA{}, $\sim$0.025 \AA{}, and $\sim$9.5 seconds, respectively. Slit-jaw images (SJI) were taken at a cadence of $\sim$19 seconds in
each of the 1400\AA{}~and 2796\AA{}~passbands. The starting exposure time was $\sim$8 seconds, and the automatic exposure control successfully
reduced the exposure time of the NUV spectra to 2.4 seconds after 17:27:09 UT. during the flare. The level 2 data are used here. The absolute
wavelength calibration and cross-channel coalignment have been done similarly to the 2014 September 6 dataset. We notice that the
Si~{\sc{iv}}~spectra obtained by IRIS at the southern part of the eastern ribbon have been analyzed by \cite{LiZ2015} for the study of slipping
magnetic reconnection. The IRIS data in this event has also been used by \cite{Cheng2015a} and \cite{LiD2015} to study the formation of flux ropes and quasi-periodic pulsations, respectively. In this paper we focus mainly on the Fe~{\sc{xxi}}~1354.08\AA{}~line emission at the northern part of the eastern ribbon. 

The AIA images have also been analyzed. Similar to the 2014 September 6 dataset, we used the AIA 1600\AA{}~images for the coalignment between
IRIS and AIA data. For the AIA data we mainly focus on the 131\AA{}~image sequence, which has a cadence of 12 seconds and a pixel size of
0.6$^{\prime\prime}$. We notice that many pixels are saturated in the 131\AA{}~images taken around the flare peak time. An online movie m3.mp4
shows the evolution of the entire flare region in the AIA 1600\AA{}~and 131\AA{}~passbands.

RHESSI data are not available for this flare. SXR images were taken with the HINODE/XRT instrument in several filters, although the field of
view (FOV) of XRT only has partial overlap with that of IRIS/SJI. XRT images taken in the Al\_thick, Be\_thin and Be\_med filters appear to be
similar. We thus use only the Al\_thick images in this paper. To coalign these images we have performed a cross-correlation of every 5 images.
These internally aligned XRT images have then been aligned with the AIA 131\AA{}~images, and thus also the IRIS images. A movie of the XRT
images is available online (m4.mp4).

Figure~\ref{fig.8}(A)-(F) presents the IRIS/SJI 1400\AA{}, AIA 131\AA{} and XRT Al\_thick images taken at two different times. A dashed line was
drawn to indicate a fast motion seen in these images. The space-time maps for this dashed line are shown in Figure~\ref{fig.8}(G)-(I).

We present the time sequence of IRIS observation (including SJI 1400\AA{} images and three spectral windows) and the corresponding AIA
131\AA{}~images in an online movie m5.mp4. The GOES 1--8\AA{} flux and XRT Al\_thick flux integrated over the box marked in
Figure~\ref{fig.8}(E)\&(H) are also shown in the movie. These two light curves appear to have a similar trend, and the general trend of their time
derivatives is also similar. We have noticed an interesting Fe~{\sc{xxi}}~1354.08\AA{}~emission feature around the slit position
117.8$^{\prime\prime}$. Since our following analysis is focused mainly on this feature, we calculated the NUV and FUV continua by averaging the
IRIS spectra over 5 pixels around slit position 117.8$^{\prime\prime}$ and wavelength ranges of 1355.08\AA{}--1355.43\AA{}~and
2794.10\AA{}--2794.45\AA{}, respectively. The time history of the Mg~{\sc{ii}}~2798.8\AA{}~intensity (integrated over
2798.65\AA{}--2799.00\AA{}, NUV continuum subtracted) was also calculated over the same slit positions. Figure~\ref{fig.9} is a snapshot of this
movie. Movie m6.mp4 shows the time evolution of the Fe~{\sc{xxi}}~1354.08\AA{}~and Si~{\sc{iv}}~1402.77\AA{}~spectra at the full-cadence
($\sim$9.5 s) in a smaller section of the slit. Similar to movie m2.mp4, different images have been scaled differently in m6.mp4 to reveal a
better contrast between the Fe~{\sc{xxi}}~emission and the other emission features.

\subsection{Results and discussion}

From the XRT Al\_thick and AIA 131\AA{}~images we can see that the sigmoid consists of two ``J"-shape loops which evolve rapidly during the
flare. The IRIS slit crosses the ribbon where the eastern ``J" loop is rooted. A fast motion of bright front can be clearly identified from
17:29:48 UT to 17:32:37 UT at the northern part of this ribbon (Figure~\ref{fig.8}). This apparent motion is seen in not only the hot XRT
Al\_thick and AIA 131\AA{}~channels, but also in the IRIS/SJI 1400\AA{}~passband which samples mainly the Si~{\sc{iv}}~lines formed at a
temperature of $\sim$10$^{4.9}$ K. The similar dynamics of both the cool and hot plasma is consistent with the scenario of chromospheric heating
and evaporation in sequentially reconnected loops, thus supporting the interpretation of slipping reconnection by \cite{LiZ2015}. The apparent
speed of the motion has been estimated to be $\sim$200 km~s$^{-1}$ from the space-time maps (Figure~\ref{fig.8}(G)-(I)), which is much higher
than the sound speed at a temperature of $\sim$10$^{4.9}$ K. We believe that the fast apparent motion is not real mass motion but just a
manifestation of the progression of the magnetic reconnection site along the flare arcade. In other words, the fast apparent motion is ultimately due to a succession of chromospheric footpoints being heated with nonthermal electrons accelerated by reconnection in a succession of loops as the magnetic reconnection site progresses along the flare arcade.

Figure~\ref{fig.9} shows clear enhancement of both the FUV and NUV continua at both the southern and northern parts of the covered ribbon. In
the southern part, the Mg~{\sc{ii}}~2798.8\AA{}~line turns from absorption to emission at about 17:24:29 UT and clear continuun enhancement can
be identified $\sim$20 s later (Movie m5.mp4). Obvious red wing enhancement frequently appears in the line profiles of C~{\sc{i}}, Si~{\sc{iv}}
and Mg~{\sc{ii}}, indicating downward motion of the chromosphere at a few tens km~s$^{-1}$ due to the heating related overpressure. However, no
obvious signature of Fe~{\sc{xxi}}~1354.08\AA{}~is found at this part, suggesting that the chromospheric heating is likely not intense enough to
produce $\sim$11 MK plasma there.

The chromospheric heating appears much more efficient in the northern part of the ribbon, where we see greatly enhanced continua and largely
blueshifted Fe~{\sc{xxi}}~1354.08\AA{}. From the online movie m5.mp4 we can see that the continuum enhancement and
Mg~{\sc{ii}}~2798.8\AA{}~emission first appears around 17:25:25 UT at slit position $\sim$126$^{\prime\prime}$. As the ribbon extends in the
north-south direction after 17:30:07 UT, the slit positions where the continua and Mg~{\sc{ii}}~2798.8\AA{}~intensities are enhanced expand and
cover the range of 110$^{\prime\prime}$--134$^{\prime\prime}$ around the flare peak time. Clear Fe~{\sc{xxi}}~1354.08\AA{}~signature can be
identified at slit position $\sim$128$^{\prime\prime}$ at 17:28:52 UT. The blueshifted Fe~{\sc{xxi}}~emission can be more easily identified as
the ribbon extends. The largest blue shift appears to be found at the southern edge of the ribbon, where the continua and
Mg~{\sc{ii}}~2798.8\AA{}~intensities are also the largest. This suggests that the largest blue shift is likely associated with the newly heated
plasma. Before the GOES SXR peak, we can clearly see that entirely blueshifted Fe~{\sc{xxi}}~emission features continuously migrate to the rest
wavelength position of the line, leading to more and more nearly stationary Fe~{\sc{xxi}}~emission at the locations of the hot flare loops. This
behavior provides support to the scenario of the hot emission in flare loops being produced through an accumulation of the evaporated plasma. A
smooth transition from the largest blue shift at the outer edge of the ribbon to a zero shift in the flare loops, which has been found by
\cite{Young2015} in another flare, is clearly present in the later stage of the impulsive phase ($\sim$17:35 UT -- 17:45 UT). The main
evaporation process appears to cease after the GOES SXR peak, when the blue shift of Fe~{\sc{xxi}}~almost disappears and the intensity of the
nearly stationary Fe~{\sc{xxi}}~emission decreases with time due to the cooling of the hot plasma.

\begin{figure*}
\centering {\includegraphics[width=\textwidth]{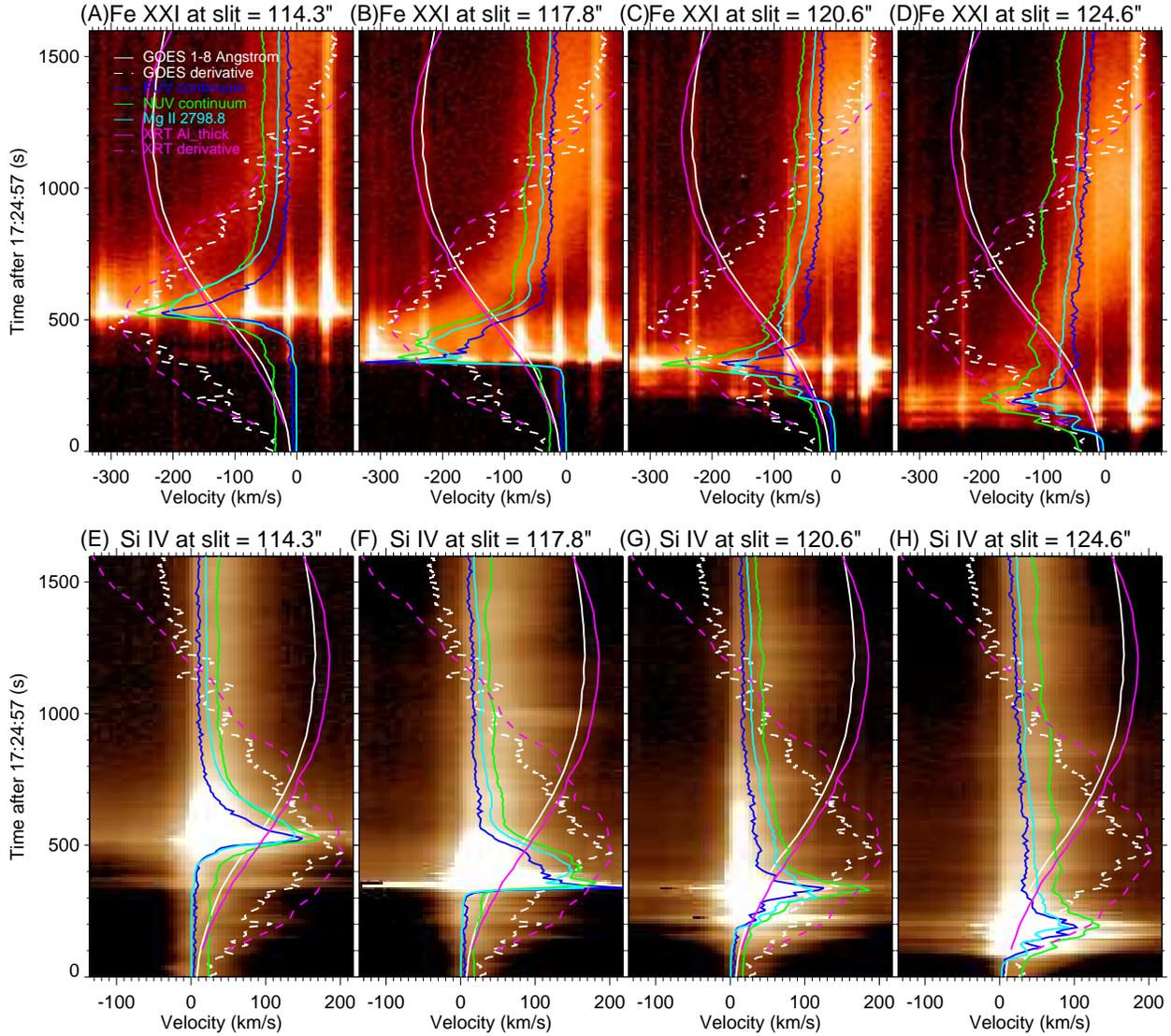}} \caption{ Time evolution of the spectral line profiles of
Fe~{\sc{xxi}}~1354.08\AA{}~and Si~{\sc{iv}}~1402.77\AA{} at four different positions on the slit in the 2014 September 10 observation. The GOES
1-8\AA{} flux (white solid) and its time derivative (white dashed), XRT Al\_thick intensity integrated over the box shown in
Figure~\ref{fig.8}(E) (purple solid) and its time derivative (purple dashed), continua (blue for FUV and green for NUV) and
Mg~{\sc{ii}}~2798.8\AA{}~(cyan) intensities at the corresponding slit positions are overplotted.} \label{fig.10}
\end{figure*}

Figure~\ref{fig.10} shows the wavelength-time plots for Fe~{\sc{xxi}}~1354.08\AA{}~and Si~{\sc{iv}}~1402.77\AA{}~at four different positions in
the northern part of the ribbon. Largely blueshifted Fe~{\sc{xxi}}~emission appears at or after the time of sudden enhancement of the FUV
continuum at all positions, again supporting the scenario of evaporation flows driven by heating in the lower chromosphere. The maximum blue
shift of Fe~{\sc{xxi}} is on the order of $\sim$250 km~s$^{-1}$ at the slit positions of 114.3$^{\prime\prime}$, 117.8$^{\prime\prime}$ and
120.6$^{\prime\prime}$. The blue shift smoothly decreases to nearly zero within $\sim$9 min, before the peak of the GOES flux, at all four
positions. Note that the appearance of continuum enhancement and Fe~{\sc{xxi}}~emission at different times at the four different positions is
just a reflection of the ribbon expansion to the south.

Although saturated around the times of maximum continuum enhancement, the Si~{\sc{iv}}~spectra presented in Figure~\ref{fig.10}(E)-(H) reveal
some results similar to those in the 2010 September 6 event. The strong correlation of Si~{\sc{iv}}~1402.77\AA{}, continua and
Mg~{\sc{ii}}~2798.8\AA{}~line intensities supports the suggestion of FUV continuum enhancement being driven by back-warming from the C~{\sc{ii}}
and Si~{\sc{iv}} lines \citep{Doyle1992}. The obvious red wing asymmetry in the impulsive phase and decay phase are likely signatures of the
chromospheric condensation \citep{Fisher1985a} and cooling of the heated plasma \citep{Brosius2003}, respectively.

\begin{figure*}
\centering {\includegraphics[width=0.8\textwidth]{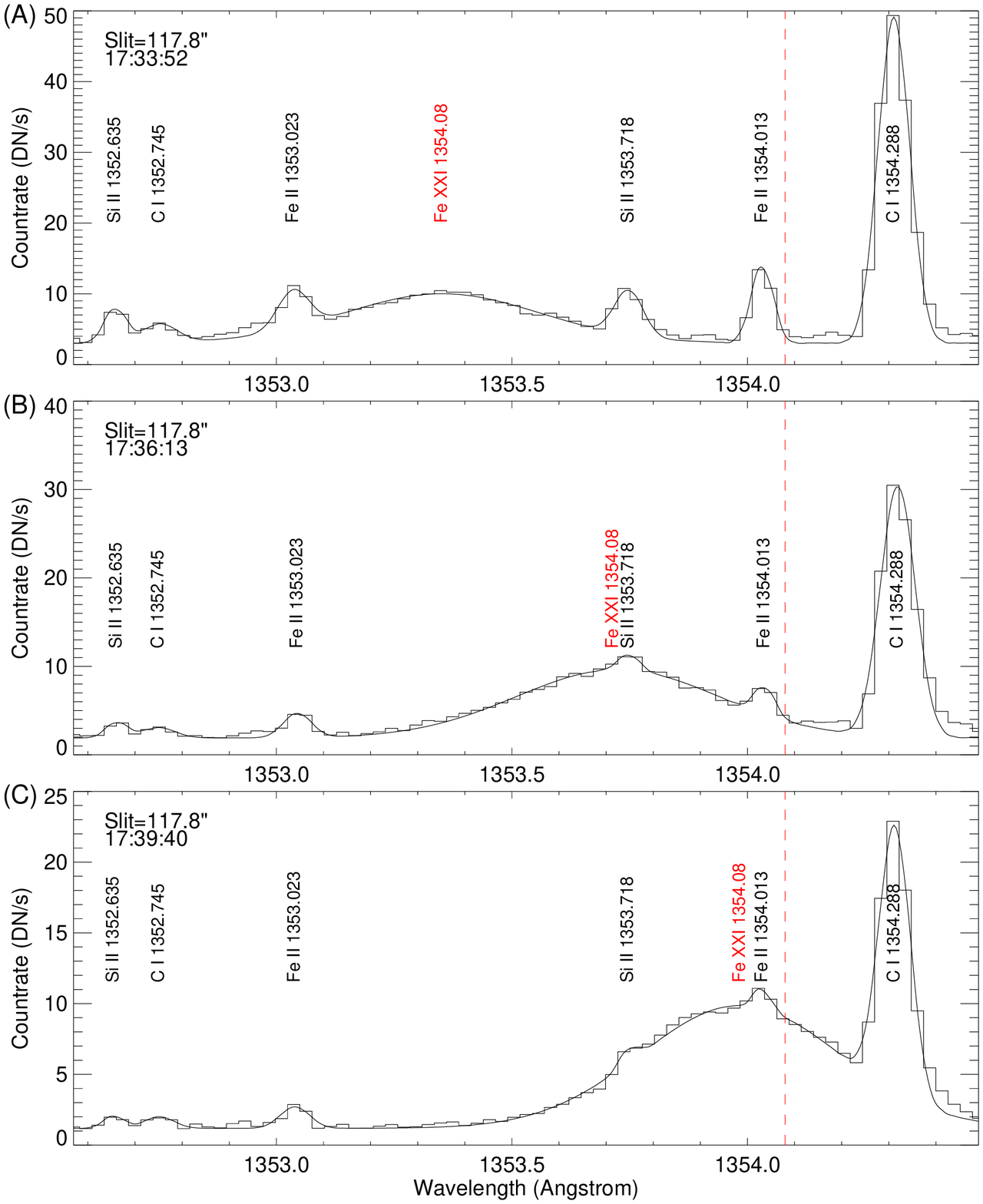}} \caption{ Three examples of multi-Gaussian fit to Fe~{\sc{xxi}}~1354.08\AA{}~and
several lines nearby in the 2014 September 10 observation. The observed and fitted line profiles are shown as the histograms and smooth curves,
respectively. The dashed line indicates the rest wavelength of Fe~{\sc{xxi}}~1354.08\AA{}. } \label{fig.11}
\end{figure*}

\begin{figure*}
\centering {\includegraphics[width=0.8\textwidth]{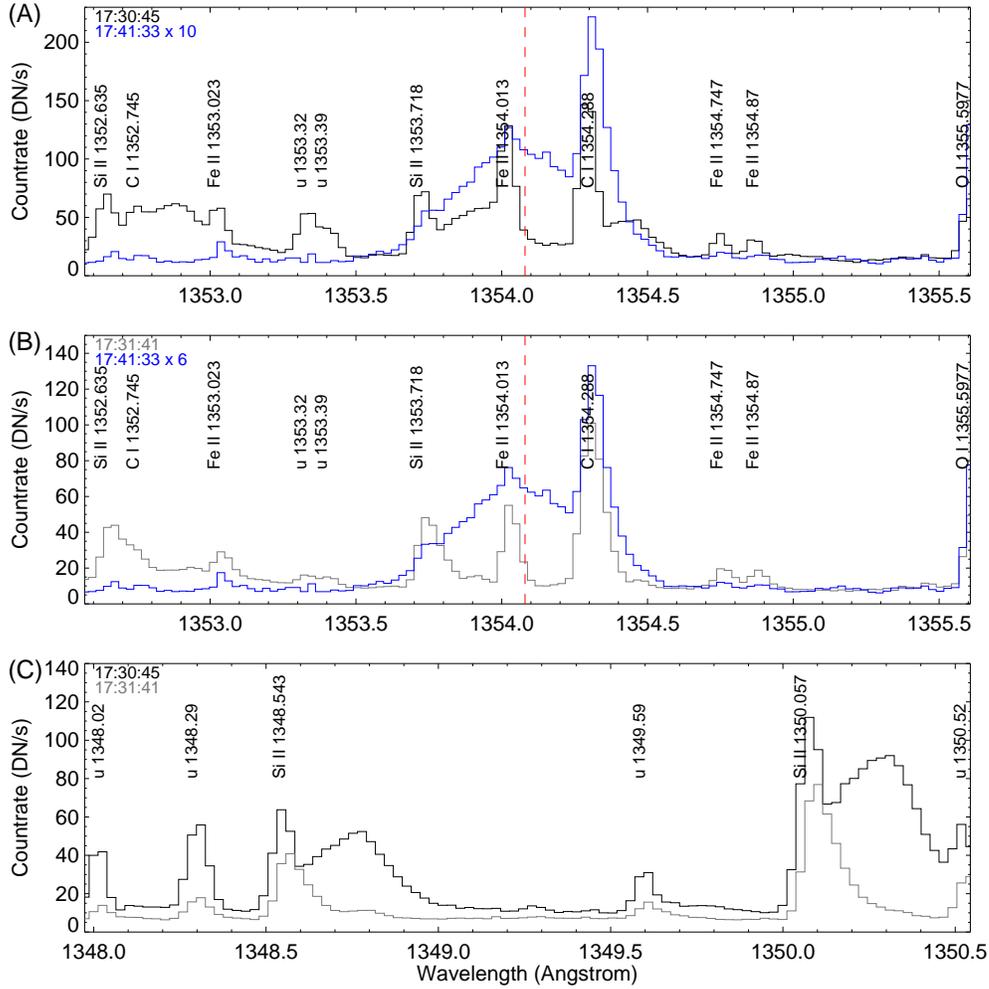}} \caption{ Line profiles of the 1354\AA{}~(A and B) and 1349\AA{}~(C) spectral
windows at 17:30:45 UT and 17:31:41 UT on 2014 September 10, at slit position 117.8$^{\prime\prime}$. As a comparison the line profile at
17:41:33 UT is re-scaled and overplotted in panels (A) and (B). The dashed line indicates the rest wavelength of Fe~{\sc{xxi}}~1354.08\AA{}.
Several previously unidentified lines are marked as the 'u' followed by the approximate wavelengths.} \label{fig.12}
\end{figure*}

Since the ribbon moves a lot in the slit direction and there are many loops overlapping with each other, it is difficult to identify and track
the Fe~{\sc{xxi}}~emission in a single loop. Fortunately, we find that the Fe~{\sc{xxi}}~emission feature at the slit position
117.8$^{\prime\prime}$ appears not moving in the slit direction from $\sim$17:32:28 UT to 17:41:23 UT. We track the time evolution of the
Fe~{\sc{xxi}}~line parameters at this position. Similar to the 2010 September 6 event, we average the line profiles over 5 pixels around slit
position 117.8$^{\prime\prime}$ and derive the Fe~{\sc{xxi}}~1354.08\AA{}~line parameters through a seven-component Gaussian fit to the
Fe~{\sc{xxi}}~1354.08\AA{}~line and six cool lines nearby. Figure~\ref{fig.11} shows three examples of multi-Gaussian fits. We notice that extra
emission around the short wavelength edge of the spectral window, similar to those seen in the 2014 September 6 flare (Figure~\ref{fig.5}), is
also clearly present in the line profiles obtained from 17:30:35 UT and 17:32:19 UT. Figure~\ref{fig.12}(A)-(B) show two examples. There the
line profile obtained at 17:41:33 UT has been re-scaled to let the line free part of the spectrum from 1355.1\AA{}--1355.4\AA{}~matches those at
17:30:45 UT and 17:31:41 UT. The extra emission may indicate a very large blue shift of Fe~{\sc{xxi}}, $\sim$300 km~s$^{-1}$ or even greater
since part of the Fe~{\sc{xxi}}~line profile may have been shifted out of the spectral window. However, as we discussed above, the presence of a
broad redshifted component (e.g., at 17:30:45 UT) or red wing enhancement (e.g., at 17:31:41 UT) in three other
Si~{\sc{ii}}~lines (1348.543\AA{}~and 1350.057\AA{}~shown in Figure~\ref{fig.12}(C), 1353.718\AA{}~shown in Figure~\ref{fig.12}(A)-(B)) suggests a significant contribution of the extra emission by the redshifted
Si~{\sc{ii}}~1352.635\AA{}~line. We decide not to perform multi-Gaussian fitting to the line profiles obtained from 17:30:35 UT and 17:32:19 UT
since it is difficult to decompose the contribution from the highly blueshifted Fe~{\sc{xxi}}~line and the redshifted Si~{\sc{ii}}~emission.

\begin{figure*}
\centering {\includegraphics[width=0.8\textwidth]{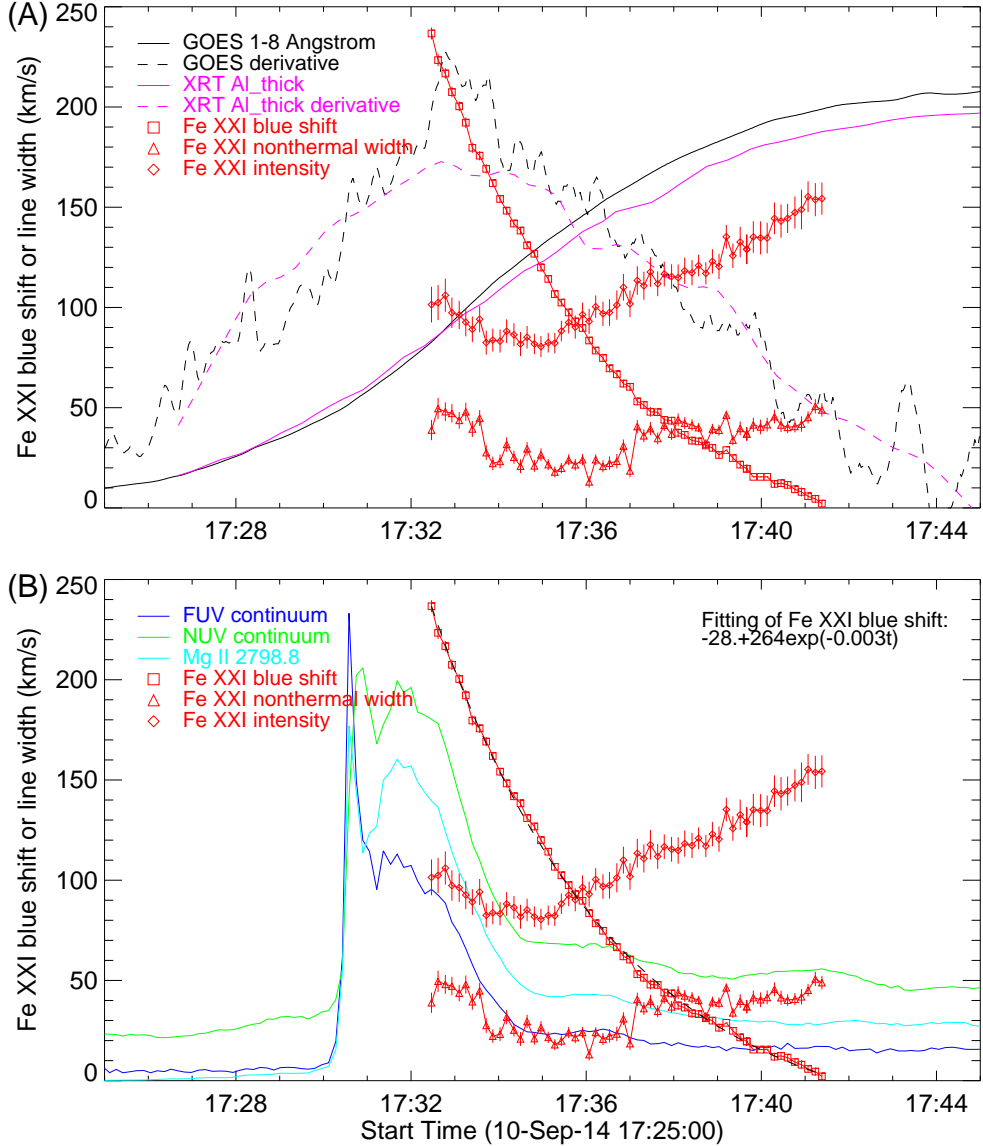}} \caption{ Time history of the line parameters of Fe~{\sc{xxi}}~1354.08\AA{}, the
GOES 1-8\AA{} flux and its time derivative, XRT Al\_thick intensity integrated over the box shown in Figure~\ref{fig.8}(E) and its time
derivative, continua and Mg~{\sc{ii}}~2798.8\AA{}~intensities averaged over 5 pixels around the slit position 117.8$^{\prime\prime}$, in the
2014 September 10 observation. The error bars represent the 1-$\sigma$ uncertainties from the multi-Gaussian fitting. Note that highly
blueshifted ($\sim$300 km~s$^{-1}$ or larger) Fe~{\sc{xxi}}~line may be present from 17:30:35 and 17:32:19 but cannot be unambiguously
identified. The black dashed line in panel (B) is the exponential fit of the time evolution of Fe~{\sc{xxi}}~blue shift. The variable t is the time (second ) after 17:32:28 UT. } \label{fig.13}
\end{figure*}

The derived Fe~{\sc{xxi}}~1354.08\AA{}~line parameters and the 1-$\sigma$ uncertainties are plotted in Figure~\ref{fig.13}. The 1/e line width
values are mostly in the range of 60--80 km~s$^{-1}$, leading to a nonthermal broadening of 10--50 km~s$^{-1}$ after subtracting the thermal and
instrumental broadening. Different from the 2014 September 6 event, the line intensity decreases slightly before increasing continuously till
the end of the measurements at 17:41:23 UT. As we explained above, Fe~{\sc{xxi}}~emission with a blue shift larger than 200 km~s$^{-1}$ first
appears in the time range of 17:30:35 -- 17:32:28 UT, same as or slightly after the time of the sudden enhancement of the continua and
Mg~{\sc{ii}}~2798.8\AA{}~intensities. The maximum blue shift appears to precede the peaks of the SXR (GOES and XRT) derivatives by
$\sim$0.5--2.0 minutes. This delay might be qualitatively consistent with a recent radiative hydrodynamic simulation (RADYN) of chromospheric evaporation \citep{RubiodaCosta2015}, where the maximum upflow speed is
achieved before the maximum of the injected electron flux, and thus HXR flux or time derivative of SXR flux. However, we have to remember that the maximum blue shift actually occurs at different times at different pixels within the ribbon \citep{Graham2015}. Note that the XRT flux is integrated over a small region around the position where the Fe~{\sc{xxi}}~1354.08\AA{}~line parameters are measured.

The blue shift smoothly decreases to $\sim$0 km~s$^{-1}$ in $\sim$9 minutes. The time evolution of the blue shift from 17:32:28 to 17:41:23
UT appears to be well fitted with an exponential function. It suggests that the evaporation process evolves smoothly, perhaps indicating a continuous
energy input. However, there is an uncertainty as to whether overlying loops corresponding to different footpoints may be contributing to the
Fe~{\sc{xxi}}~1354.08\AA{}~emission feature at slit position 117.8$^{\prime\prime}$. If this was the case, then we would expect very broad
Fe~{\sc{xxi}} line profiles or perhaps multiple components which does not appear to be the case.

\begin{figure*}
\centering {\includegraphics[width=0.8\textwidth]{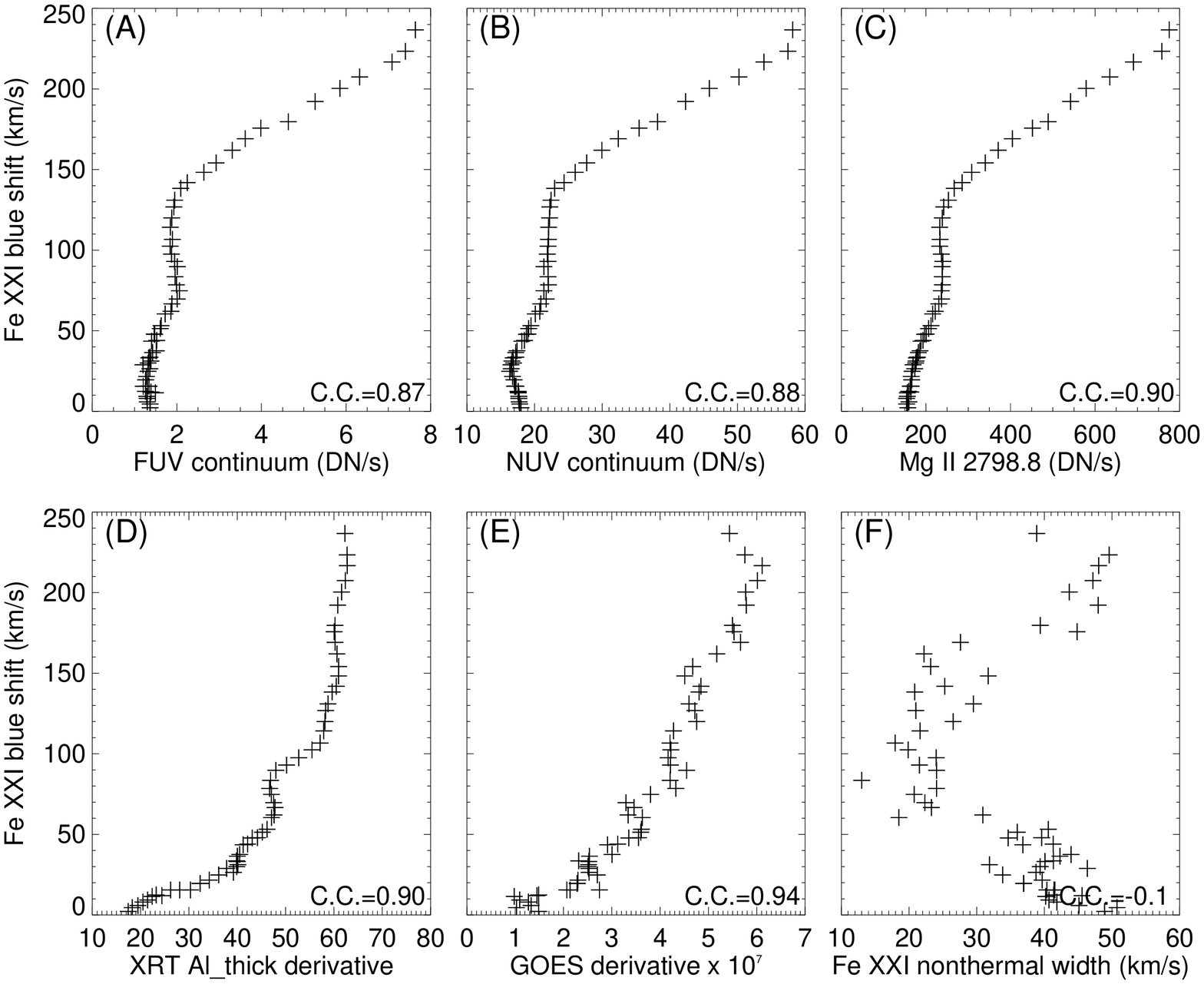}} \caption{ Scatter plots showing the correlation between Fe~{\sc{xxi}}~blue shift and
other parameters in the X1.6 flare on 2014 September 10. The linear Pearson correlation coefficient (C.C.) is marked in each panel. }
\label{fig.14}
\end{figure*}

Figure~\ref{fig.14} presents various scatter plots for the relationship between Fe~{\sc{xxi}}~blue shift and other measured quantities. As
expected from the scenario of evaporation flows being driven by chromospheric heating, larger blue shifts are found when the intensities of
Mg~{\sc{ii}}~2798.8\AA{} and the FUV and NUV continua are larger. The correlation appears to be linear for blue shifts larger than $\sim$130
km~s$^{-1}$. The chromospheric intensities do not drop too much when the Fe~{\sc{xxi}}~blue shift reaches smaller values, simply because they
drop faster than the decrease of the blue shift. Similar to the 2014 September 6 event, here we also find a good correlation between the
Fe~{\sc{xxi}}~blue shift and the SXR derivative. The best correlation is still the one between the Fe~{\sc{xxi}}~blue shift and the derivative
of GOES, which has a linear correlation coefficient of 0.94. As we explained above, this correlation means that the evaporation flow velocity
largely depends on the energy deposition rate, and thus the heating rate.

The scatter plot for the relationship between the blue shift and nonthermal width (Figure~\ref{fig.14}(F)) appears surprising. We see a positive
correlation when the blue shift is larger than $\sim$100 km~s$^{-1}$ and a negative correlation when the blue shift is smaller than $\sim$100
km~s$^{-1}$. More case studies will tell how common this characteristic is.

\section{Summary}
Using IRIS observations, we have tracked the complete evolution of the hot evaporation flows continuously in the M1.1 flare occurring on 2014
September 6 and the X1.6 flare on 2014 September 10. For both flares, the hot evaporation flows marked by the blueshifted
Fe~{\sc{xxi}}~1354.08\AA{}~emission are tracked with a cadence higher than 10 seconds. The relationship between chromospheric evaporation and
heating, as indicated by the enhancement of the UV continuum and Mg~{\sc{ii}}~2798.8\AA{} line emission, has been studied as well. We have also
investigated the correlation between the evaporation flow velocity and the energy deposition rate, as represented by the HXR flux measured by
RHESSI or the time derivative of the SXR flux observed by GOES and HINODE/XRT. The main results are summarized in the following.

(1) The NUV spectra at the ribbons of both flares are dominated by emission lines. The is true for wavelengths at least between
2791\AA{}--2834\AA{} and indicates strong heating in both the chromosphere and photosphere.

(2) The Fe~{\sc{xxi}}~line is completely blueshifted at or near the flare ribbons indicated by the enhancement of the UV continuum. The absence
of the rest component, which was often observed in flare lines by previous instruments, agrees with simulations of chromospheric evaporation in
a single flare loop \citep[e.g.,][]{Emslie1987,Liu2009} and suggests that the resolution of IRIS is high enough to separate the evaporation
flows from the surrounding hot plasma in flare loops.

(3) The flare loops have little Fe~{\sc{xxi}}~emission in the early stage of the impulsive phase. Fe~{\sc{xxi}}~emission appears to shift from
the footpoints to the higher parts of the flare loops in the later stage of the impulsive phase, until the entire flare loop emits in
Fe~{\sc{xxi}}~at the time of the peak in the GOES flux. This behavior provides support to the scenario of the SXR emission in flare loops being
produced through an accumulation of the evaporated hot plasma from the chromosphere, although we can not rule out the possibility of certain
degrees of in-situ heating in the corona.

(4) The maximum blue shift of Fe~{\sc{xxi}}~is on the order of 200~km~s$^{-1}$, though possible signatures of $\sim$300~km~s$^{-1}$ or larger
blue shifts have also been identified. The maximum blue shift appears approximately at the same time as or slightly after the impulsive
enhancement of the FUV/NUV continuum and Mg~{\sc{ii}}~2798.8\AA{} emission, indicating that the hot evaporation flow is closely related to
intense heating of the lower chromosphere.

(5) After reaching the maximum, the blue shift decreases smoothly to almost stationary within a few minutes. The smooth decrease may suggest
that the energy input into the lower atmosphere is smooth and continuous in these flares. The velocity decrease appears to be exponential in
time, especially for the X1.6 flare.

(6) The Fe~{\sc{xxi}}~blue shift appears to correlate with the hard X-Ray flux (or derivative of soft X-Ray flux), which is in general agreement
with the picture of chromospheric evaporation driven by nonthermal electron heating. A nearly linear correlation between the blue shift and GOES
derivative is found in both flares, suggesting that the evaporation flow velocity largely depends on the rate of energy deposition or heating.

(7) The intensities of the FUV continuum, NUV continuum, and the chromospheric and transition region lines appear to be correlated. This
correlation supports the suggestion of FUV continuum enhancement in flares being driven by back-warming from the C~{\sc{ii}} and Si~{\sc{iv}}
lines \citep{Doyle1992}. The decay of these intensities is faster than the decrease of the Fe~{\sc{xxi}}~blue shift.

(8) The chromospheric and transition region lines often show obvious enhancement in the red wing of the line profiles at the flare ribbons.
Several Si~{\sc{ii}}~lines sometimes even show a distinct broad and redshifted component. The red wing enhancement and redshifted component
suggest that the speed of chromospheric condensation might be larger (up to $\sim$100 km~s$^{-1}$) than previously found from hydrodynamic simulations and from single Gaussian fits to the cool lines of EIS and CDS. Their presence also complicates the detection of highly blueshifted
Fe~{\sc{xxi}}~1354.08\AA{} around the time of maximum continuum enhancement. The red wing enhancement is found in not only the impulsive phase
but also the decay phase. The latter is likely caused by cooling of the heated plasma in the corona.

(9) In the X1.6 flare, we see a bright intensity front propagating with an apparent speed of $\sim$200 km~s$^{-1}$ along the ribbon in IRIS/SJI
1400\AA{}, XRT Al\_thick and AIA 131\AA{}~images. This fast apparent motion is not real mass motion but just a reflection of the chromospheric
heating and evaporation in sequentially reconnected loops.

\begin{acknowledgements}
IRIS is a NASA small explorer mission developed and operated by LMSAL with mission operations executed at NASA Ames Research center and major
contributions to downlink communications funded by the Norwegian Space Center (NSC, Norway) through an ESA PRODEX contract. Hinode is a Japanese
mission developed and launched by ISAS/JAXA, with NAOJ as domestic partner and NASA and STFC (UK) as international partners. It is operated by
these agencies in co-operation with ESA and the NSC (Norway). This work is supported by NASA contracts 8100002705, NNM07AB07C and SP02H1701R
from LMSAL to SAO, NASA grants NNX13AE06G and NNX15AJ93G, and NSF grant AGS-1159353. H.T. thanks Fatima Rubio da Costa and Xin Cheng for
insightful discussion.
\end{acknowledgements}


\begin{thebibliography}{}
\bibitem[Antonucci et al.(1982)]{Antonucci1982}
Antonucci, E., Gabriel, A. H., Acton, L. W., et al. 1982, SoPh, 78, 107
%
\bibitem[Antiochos et al.(1999)]{Antiochos1999}
Antiochos, S., Devore, C. R., \& Klimchuk, J. A. 1999, ApJ, 510, 485
%
\bibitem[Battaglia et al.(2009)]{Battaglia2009}
Battaglia, M., Fletcher, L., Benz, A. O. 2009, ApJ, 498, 891
%
\bibitem[Boerner et al.(2012)]{Boerner2012}
Boerner, P., et al. 2012, SoPh, 275, 41
%
\bibitem[Brosius \& Holman(2010)]{Brosius2010}
Brosius, J. W., \& Holman, G. D. 2010, ApJ, 720, 1472
%
\bibitem[Brosius \& Holman(2012)]{Brosius2012}
Brosius, J. W., \& Holman, G. D. 2014, A\&A, 540, A24
%
\bibitem[Brosius(2003)]{Brosius2003}
Brosius, J. W. 2003, ApJ, 586, 1417
%
\bibitem[Brosius \& Phillips(2004)]{Brosius2004}
Brosius, J. W., \& Phillips, K. J. H. 2004, ApJ, 613, 580
%
\bibitem[Brosius(2013)]{Brosius2013}
Brosius, J. W. 2013, ApJ, 762, 133
%
\bibitem[Cassak et al.(2013)]{Cassak2013}
Cassak, P. A., Drake, J. F., Gosling, J. T., et al. 2013, ApJ, 775, L14
%
\bibitem[Carmichael(1964)]{Carmichael1964}
Carmichael, H. 1964, in AAS-NASA Symposium on Solar Flares, ed. W. N. Hess (NASA SP-50), 451
%
\bibitem[Chen \& Ding(2010)]{Chen2010}
Chen, F., Ding, M. D. 2010, ApJ, 724, 640
%
\bibitem[Cheng et al.(2015a)]{Cheng2015a}
Cheng, X., Ding, M. D., \& Fang, C. 2015a, ApJ, 804, 82
%
\bibitem[Cheng et al.(2015b)]{Cheng2015b}
Cheng, X., Hao, Q., Ding, M. D., et al. 2015b, ApJ, in press
%
\bibitem[Cohen(1981)]{Cohen1981}
Cohen, L. 1981, NASA Publ. 1069
%
\bibitem[Culhane et al.(2007)]{Culhane2007}
Culhane, J. L., et al. 2007, Sol. Phys., 243, 19
%
\bibitem[Curdt et al.(2001)]{Curdt2001}
Curdt, W., Brekke, P., Feldman, U., Wilhelm, K., Dwivedi, B.N., Sch\"uhle, U., \& Lemaire, P. 2001, A\&A, 375, 591
%
\bibitem[Curdt et al.(2004)]{Curdt2004}
Curdt, W., Landi, E., \& Feldman, U. 2004, A\&A,427,1045
%
\bibitem[Czaykowska et al.(1999)]{Czaykowska1999}
Czaykowska, A., et al. 1999, ApJ, 521, L75
%
\bibitem[Czaykowska et al.(2001)]{Czaykowska2001}
Czaykowska, A., et al. 2001, ApJ, 552, 849
%
\bibitem[Del Zanna et al.(2006)]{DelZanna2006}
Del Zanna, G., Berlicki, A., Schmieder, B., \& Mason, H. E. 2006, SoPh, 234, 95
%
\bibitem[Del Zanna et al.(2011)]{DelZanna2011}
Del Zanna, G., O$^{\prime}$Dwyer, B., Mason, H. E. 2011, A\&A, 535, A46
%
\bibitem[Dennis \& Zarro(1993)]{Dennis1993}
Dennis, B. R., \& Zarro, D. M. 1993, Sol. Phys., 146, 177
%
\bibitem[De Pontieu et al.(2014)]{DePontieu2014}
De Pontieu, B., et al. 2014, Sol. Phys., 289, 2733
%
\bibitem[Doschek et al.(1980)]{Doschek1980}
Doschek, G. A., Feldman, U., Kreplin, R. W., \& Cohen, L. 1980, ApJ, 239, 725
%
\bibitem[Doschek et al.(2013)]{Doschek2013}
Doschek, G. A., Warren, H. P., Young, P. R. 2013, ApJ, 767, 55
%
\bibitem[Doyle \& Phillips(1992)]{Doyle1992}
Doyle, J. G., Phillips, K. J. H. 1992, A\&A, 257, 773
%
\bibitem[Ding et al.(1995)]{Ding1995}
Ding, M.-D., Fang, C., \& Huang, Y.-R. 1995, Sol. Phys., 158, 81
%
\bibitem[Ekberg \& Feldman(2003)]{Ekberg2003}
Ekberg, J. O., \& Feldman, U. 2003, ApJS, 148, 567
%
\bibitem[Emslie \& Alexander(1987)]{Emslie1987}
Emslie, G. A., Alexander, D. 1987, Sol. Phys. 110, 295
%
\bibitem[Falchi et al.(2006)]{Falchi2006}
Falchi, A., Teriaca, L., Maltagliati, L. 2006, Sol. Phys. 239, 193
%
\bibitem[Feldman et al.(1980)]{Feldman1980}
Feldman, U., Doschek, G. A., Kreplin, R. W., \& Mariska, J. T. 1980, ApJ, 241, 1175
%
\bibitem[Feldman et al.(1997)]{Feldman1997}
Feldman, U., Behring, W. E., Curdt, W. et al. 1997, ApJS, 113, 195
%
\bibitem[Feldman et al.(2000)]{Feldman2000}
Feldman, U., Curdt, W., Landi, E., \& Wilhelm, K. 2000, ApJ, 544, 508
%
\bibitem[Fisher et al.(1985a)]{Fisher1985a}
Fisher, G. H., Canfield, R. C., McClymont, A. N. 1985a, ApJ, 289, 414
%
\bibitem[Fisher et al.(1985b)]{Fisher1985b}
Fisher, G. H., Canfield, R. C., McClymont, A. N. 1985b, ApJ, 289, 425
%
\bibitem[Fisher et al.(1989)]{Fisher1989}
Fisher, G. H. 1989, ApJ, 346, 1019
%
\bibitem[Golub et al.(2007)]{Golub2007}
Golub, L., DeLuca, E., Austin, G., et al. 2007, Solar Phys, 243, 63
%
\bibitem[Graham et al.(2011)]{Graham2011}
Graham, D. R., Fletcher, L., \& Hannah, I. G. 2011, A\&A, 532, A27
%
\bibitem[Graham \& Gauzzi(2015)]{Graham2015}
Graham, D. R., Gauzzi, G. 2015, ApJL, 807, L22
%
\bibitem[Harra et al.(2005)]{Harra2005}
Harra, L. K., D\'{e}moulin, P., Mandrini, C. H., et al. 2005, A\&A, 438, 1099
%
\bibitem[Harrison et al.(1995)]{Harrison1995}
Harrison, R. A., et al. 1995, Sol. Phys., 162, 233
%
\bibitem[Heinzel \& Kleint(2014)]{Heinzel2014}
Heinzel, P., Kleint, L. 2014, ApJL, 794, L23
%
\bibitem[Hirayama(1974)]{Hirayama1974}
Hirayama, T. 1974, Sol. Phys., 34, 323
%
\bibitem[Hudson(1991)]{Hudson1991}
Hudson, H. S. 1991, BAAS, 23, 1064
%
\bibitem[Kelly(1979)]{Kelly1979}
Kelly, R. L. 1979, Atomic emission lines in the near ultraviolet: hydrogen through krypton. Greenbelt, MD: Goddard Space Flight Center.
%
\bibitem[Kelly(1987)]{Kelly1987}
Kelly, R. L. 1987, J. Phys. Chem. Ref. Data 16, Suppl. 1
%
\bibitem[Kopp \& Pneuman(1976)]{Kopp1976}
Kopp, R. A., \& Pneuman, G. W. 1976, Sol. Phys., 50, 85
%
\bibitem[Krucker et al.(2015)]{Krucker2015}
Krucker, S., Saint-Hilaire, P., Hudson, H. S., et al. 2015, ApJ, 802, 19
%
\bibitem[Landi et al.(2013)]{Landi2013}
Landi, E., Young, P. R., Dere, K. P., Del Zanna, G., \& Mason, H. E. 2013, ApJ, 763, 86
%
\bibitem[Lemen et al.(2012)]{Lemen2012}
Lemen, J. R., et al. 2012, Solar Phys., 275, 17
%
\bibitem[Li et al.(2015)]{LiD2015}
Li, D., Ning, Z. J., Zhang, Q. M. 2015, ApJ, 807, 72
%
\bibitem[Li \& Ding(2011)]{Li2011}
Li, Y., Ding, M. D. 2011, ApJ, 727, 98
%
\bibitem[Li et al.(2015)]{Li2015}
Li, Y., Ding, M. D., Qiu, J., Cheng, J. X. 2015, submitted to ApJ
%
\bibitem[Li \& Zhang(2015)]{LiZ2015}
Li, T., Zhang, J. 2015, ApJL, 804, L8
%
\bibitem[Lin et al.(2002)]{Lin2002}
Lin, R. P., et al. 2002, Sol. Phys., 210, 3
%
\bibitem[Liu et al.(2013)]{Liu2013}
Liu, W.-J., Qiu, J., Longcope, D. W., Caspi, A. 2013, ApJ, 770, 111
%
\bibitem[Liu et al.(2006)]{Liu2006}
Liu, W., et al. 2006, ApJ, 649, 1124
%
\bibitem[Liu et al.(2009a)]{Liu2009}
Liu, W., Petrosian, V., Mariska, J. T. 2009a, ApJ, 702, 1553
%
\bibitem[Liu et al.(2009b)]{Liu2009b}
Liu, W., Petrosian, V., Dennis, B. R., Holman, G. D. 2009b, ApJ, 693, 847
%
%
\bibitem[Mason et al.(1986)]{Mason1986}
Mason, H. E., Shine, R. A., Gurman, J. B., Harrison, R. A. 1986, ApJ, 309, 435
%
\bibitem[Mariska \& Doschek(1993)]{Mariska1993}
Mariska, J. T., \& Doschek, G. A. 1993, ApJ, 419, 418
%
\bibitem[Milligan et al.(2006a)]{Milligan2006a}
Milligan,R.O., Gallagher, P. T., Mathioudakis, M.,\& Keenan, F. P. 2006a, ApJL, 642, L169
%
\bibitem[Milligan et al.(2006b)]{Milligan2006b}
Milligan, R. O., Gallagher, P. T., Mathioudakis, M., et al. 2006b, ApJL, 638, L117
%
%
\bibitem[Milligan \& Dennis(2009)]{Milligan2009}
Milligan, R. O., Dennis, B. R. 2009, ApJ, 699, 968
%
\bibitem[Milligan(2011)]{Milligan2011}
Milligan, R. O. 2011, ApJ, 740, 70
%
\bibitem[Neupert(1968)]{Neupert1968}
Neupert, W. M. 1968, ApJ, 153, L59
%
\bibitem[Ning et al.(2009)]{Ning2009}
Ning, Z.-J., Cao, W.-D., Huang, J. 2009, ApJ, 699, 15
%
\bibitem[Nitta et al.(2012)]{Nitta2012}
Nitta, S., Imada, S., Yamamoto, T. T. 2012, Sol. Phys., 276, 183
%
\bibitem[O$^{\prime}$Dwyer et al.(2010)]{ODwyer2010}
O$^{\prime}$Dwyer, B., Del Zanna, G., Mason, H. E., Weber, M. A., \& Tripathi, D. 2010, A\&A, 521, A21
%
\bibitem[{{Pereira} {et~al.}(2015){Pereira}, {Carlsson}, {De
  Pontieu}, \& {Hansteen}}]{Pereira2015}
{Pereira}, T.~M.~D.,  {Carlsson}, M., {De Pontieu}, B., \& {Hansteen}, V.,
 2015, submitted to \apj
 %
 %
\bibitem[Polito et al.(2015)]{Polito2015}
Polito, V., Reeves, K. K., Del Zanna, G., Golub, L., Mason, H. E. 2015, ApJ, 803, 84
%
\bibitem[Raftery et al.(2009)]{Raftery2009}
Raftery, C. L., et al. 2009, A\&A, 494, 1127
%
\bibitem[Pesnell et al.(2012)]{Pesnell2012}
Pesnell, W. D., Thompson, B. J., Chamberlin, P. C. 2012, Sol. Phys., 275, 3
%
\bibitem[Reeves et al.(2007)]{Reeves2007}
Reeves, K. K., Warren, H. P., Forbes, T. G. 2007, ApJ, 668, 1210
%
\bibitem[Rubio da Costa et al.(2015)]{RubiodaCosta2015}
Rubio da Costa, F., Petrosian, V., Liu, W., \& Carlsson, M. 2015, to be submitted to ApJ
%
\bibitem[Sandlin et al.(1977)]{Sandlin1977}
Sandlin, G. D., Brueckner, G. E., \& Tousey, R. 1977, ApJ, 214, 898
%
\bibitem[Sandlin et al.(1986)]{Sandlin1986}
Sandlin, G. D., Bartoe, J.-D. F., Brueckner, G. E., Tousey, R., Vanhoosier, M. E. 1986, ApJS, 61, 801
%
\bibitem[Sadykov et al.(2015)]{Sadykov2015}
Sadykov, V. M., Dominguez, S. V., Kosovichev, A. G., et al. 2015, ApJ, 805, 167
%
\bibitem[Silva et al.(1997)]{Silva1997}
Silva, A. V. R., Wang, H., Gary, D. E., Nitta, N., \& Zirin, H. 1997, ApJ, 481, 978
%
\bibitem[Sturrock(1966)]{Sturrock1966}
Sturrock, P. A. 1966, Nature, 211, 695
%
\bibitem[Teriaca et al.(2003)]{Teriaca2003}
Teriaca, L., Falchi, A., Cauzzi, G., et al. 2003, ApJ, 588, 596
%
\bibitem[Teriaca et al.(2006)]{Teriaca2006}
Teriaca, L., Falchi, A., Falciani, R., Cauzzi, G., \& Maltagliati, L. 2006, A\&A, 455, 1123
%
\bibitem[Tian et al.(2009)]{Tian2009}
Tian, H., Curdt, W., Teriaca, L., Landi, E., Marsch, E. 2009, A\&A 505, 307
%
\bibitem[Tian et al.(2014a)]{Tian2014}
Tian, H., Li, G., Reeves, K. K., et al. 2014a, ApJL, 797, L14
%
\bibitem[Tian et al.(2014b)]{Tian2014b}
Tian, H., DeLuca, E. E., Cranmer, S. R., et al. 2014b, Science, 346, 1255711
%
%
\bibitem[Warren \& Doschek(2005)]{Warren2005}
Warren, H. P., \& Doschek, G. A. 2005, ApJ, 618, L157
%
\bibitem[Watanabe et al.(2010)]{Watanabe2010}
Watanabe, T., Hara, H., Sterling, A. C., \& Harra, L. K. 2010, ApJ, 719, 213
%
%
\bibitem[Veronig et al.(2005)]{Veronig2005}
Veronig, A. M., Brown, J. C., Dennis, B. R., et al. 2005, ApJ, 621, 482
%
\bibitem[Xu et al.(2004)]{Xu2004}
Xu, Y., Cao, W.-D., Liu, C., et al. 2004, ApJ, 607, L131
%
%
\bibitem[Young et al.(2013)]{Young2013}
Young, P. R., Doschek, G. A., Warren, H. P., Hara, H. 2013, ApJ, 766, 127
%
\bibitem[Young et al.(2015)]{Young2015}
Young, P., Tian, H., Jaeggli, S. 2015, ApJ, 799, 218
%
\bibitem[Zarro \& Lemen(1988)]{Zarro1988}
Zarro, D. M., Lemen, J. R. 1988, ApJ, 329, 456
%
\bibitem[Zhang \& Ji(2013)]{Zhang2013}
Zhang, Q.-M., Ji, H.-S. 2013, A\&A, 557, L5
%
\end{thebibliography}
\end{document}